\documentclass[fleqn,usenatbib]{mnras}
\usepackage{newtxtext,newtxmath}
\usepackage[T1]{fontenc}
\usepackage{graphicx}
\usepackage{amsmath}
\usepackage{amssymb}
\usepackage{ulem}
\usepackage{lmodern}

\title[Discovery of a strong lensed system with MUSE]{Serendipitous discovery of a strong-lensed galaxy in integral field spectroscopy from MUSE}

\author[Galbany et al.]{Llu\'is Galbany$^{1}$\thanks{E-mail: llgalbany@pitt.edu (LG)},
Thomas E. Collett$^{2}$,
Jairo M\'endez-Abreu$^{3,4}$,
\newauthor
Sebasti\'an F. S\'anchez$^{5}$,
Joseph P. Anderson$^{6}$,
Hanindyo Kuncarayakti$^{7,8}$.
\\
$^{1}$PITT PACC, Department of Physics and Astronomy, University of Pittsburgh, Pittsburgh, PA 15260, USA.\\
$^{2}$Institute of Cosmology \& Gravitation, University of Portsmouth, Dennis Sciama Building, Portsmouth, PO1 3FX, UK.\\
$^{3}$Instituto de Astrof\'isica de Canarias, C/ V\'ia L\'actea s/n, E-38205, La Laguna, Spain.\\
$^{4}$Departamento de Astrof\'isica, Universidad de La Laguna, E-38206, La Laguna, Spain.\\
$^{5}$Instituto de Astronom\'ia, Universidad Nacional Aut\'onoma de M\'exico, A.P. 70-264, 04510 M\'exico, D.F., Mexico.\\
$^{6}$European Southern Observatory, Alonso de Cordova 3107 Casilla 19001 $-$ Vitacura $-$ Santiago, Chile.\\
$^{7}$Finnish Centre for Astronomy with ESO (FINCA), University of Turku, V\"ais\"al\"antie 20, 21500 Piikki\"o, Finland.\\
$^{8}$Tuorla Observatory, Department of Physics and Astronomy, University of Turku, V\"ais\"al\"antie 20, 21500 Piikki\"o, Finland.
}
\date{Accepted ---. Received ---; in original form ---}
\pubyear{2018}

\begin{document}
\label{firstpage}
\pagerange{\pageref{firstpage}--\pageref{lastpage}}
\maketitle

\begin{abstract}
2MASX J04035024-0239275 is a bright red elliptical galaxy at redshift 0.0661 that presents two extended sources at 2\arcsec~to the north-east and 1\arcsec~to the south-west. The sizes and surface brightnesses of the two blue sources are consistent with a gravitationally-lensed background galaxy. In this paper we present MUSE observations of this galaxy from the All-weather MUse Supernova Integral-field Nearby Galaxies (AMUSING) survey, and report the discovery of a background lensed galaxy at redshift 0.1915, together with other 15 background galaxies at redshifts ranging from 0.09 to 0.9, that are not multiply imaged. We have extracted aperture spectra of the lens and all the sources and fit the stellar continuum with STARLIGHT to estimate their stellar and emission line properties.
A trace of past merger and active nucleus activity is found in the lensing galaxy, while the background lensed galaxy is found to be star-forming. Modeling the lensing potential with a singular isothermal ellipsoid, we find an Einstein radius of 1\farcs45$\pm$0\farcs04, which corresponds to 1.9 kpc at the redshift of the lens and it is much smaller than its effective radius ($r_{\rm eff}\sim$ 9\arcsec). Comparing the Einstein mass and the STARLIGHT stellar mass within the same aperture yields a dark matter fraction of $18 \% \pm 8$ \% within the Einstein radius. The advent of large surveys such as the Large Synoptic Survey Telescope (LSST) will discover a number of strong-lensed systems, and here we demonstrate how wide-field integral field spectroscopy offers an excellent approach to study them and to precisely model lensing effects.
\end{abstract}


%%%%%%%%%%%%%%%%%%%%%%%%%%%%%%%%%%%%%%%%%%%%%%%%%%

\section{Introduction} \label{sec:intro}

General relativity predicts that light from distant galaxies is deflected by foreground massive objects, such as more nearby galaxies or clusters. In cases where the source and deflector are sufficiently well aligned, multiple images of the source will form around the lensing foreground object \citep{1915SPAW.......844E,1937PhRv...51..290Z}. 

Strong gravitational lensing produces highly magnified images that allow the study of fainter and smaller galaxies than would otherwise be possible with current instrumentation \citep{1996astro.ph..6001N}.
Also, it provides the most precise and accurate method to determine the masses of galaxies in a model-free fashion being independent of assumptions such as the initial mass function (IMF) or galaxy dynamics \citep{2010ARA&A..48...87T}.
In addition, the comparison between the inferred total mass distribution in strong lenses to their observed brightness profile provides valuable information about the galaxy dark matter content and distribution \citep{1994A&A...292..381B,2015ApJ...804L..21C}.  
Strong lensing systems can also place constraints on cosmological parameters \citep{bonvin2017, 2014MNRAS.443..969C}.

Low redshift lenses are particularly interesting systems as their Einstein radii form at a much smaller physical scale than for higher redshift lenses. This enables studies of the central matter distribution in these lenses. Their much lower distances also allow more detailed follow-up of the lens kinematics and stellar populations. Recent studies have combined these approaches to place tight constraints on the IMF \citep{2015MNRAS.449.3441S, 2017ApJ...845..157N}, the validity of General relativity (Collett 2018a), and the Hubble constant (Collett et  al 2018b, in prep). %\textbf{[HK: if any, add more pertinent references in addition to those in prep/subm]}.
Unfortunately, despite the utility of low redshift strong lenses only a handful are known \citep{2005ApJ...625L.103S, 2015MNRAS.449.3441S} due to the rarity of massive galaxies and the small volume at very low redshifts.

While some lensed systems have been discovered serendipitously by visually checking images of the lens, there have been several dedicated attempts to identify strong gravitational lensed objects around massive galaxies. 
Previous search strategies typically used both photometric or spectroscopic methods. 
Photometric methods include searching for arc morphologies (e.g. \citealt{2004ApJ...602L...9B}), or outliers in color diagrams (e.g. \citealt{2014ApJ...785..144G}), given that foreground and background objects have different colors.
The majority of these methods are based on a priori knowledge of galaxy properties, and visual inspection is usually required. Most recently, new automatic methods based on deep learning are flowering \citep{2018MNRAS.473.3895L}.
Alternatively, spectroscopic methods that are based on detecting rogue emission (or absorption) lines in galaxy spectra \citep{2006ApJ...638..703B} have been employed.

Recently, \cite{2017MNRAS.464L..46S} reported the discovery of a lensed galaxy in integral field spectroscopy (IFS) observations of SDSS J170124.01+372258.0 from the MaNGA survey \citep{Bundy15}
This methodology has been greatly extended by \citet{talbot2018} to discover 38 lenses from MaNGA, with lens redshifts spanning $0.02 < z_l < 0.13$.
Despite the clear detection of  weak emission lines in the datacube corresponding to a star-forming background galaxy at different redshift than the lens, the low spatial resolution of MaNGA data does not yet allow for a detailed understanding of the observed lensing mass distribution.

We here report the discovery of strong gravitational lensing by the galaxy 2MASX J04035024-0239275\footnote{We note that this lens has also been independently discovered by \citet{collier2018}, using the same MUSE data from our ESO programme 098.D-0115(A); PI: Galbany.}, in IFS obtained with the Multi Unit Spectroscopic Explorer (MUSE) on the Very Large Telescope (VLT), an IFU instrument with  high resolutions in both spectral ($R \sim 3000$) and spatial (0.2\arcsec/spaxel) dimensions, which permits a detailed reconstruction of the source.

In section \ref{sec:obs} we describe the observations. Then, in section \ref{sec:lens} we describe the lens model and the residual datacube after subtraction.
In section \ref{sec:source} the lensed source. In Section \ref{sec:lensmodelling} we present a simple model of the observed image configuration and the lensing mass. In section \ref{sec:conc} we summarize and list our conclusions. Throughout this work we use the best fit cosmological parameters of \cite{2016A&A...594A..13P}.

%%%%%%%%%%%%%%%%%%%%%%%%%%%%%%%%%%%%%%%%%%%%%%%%%%%%%%%
%%%%%%%%%%%%%%%%%%%%%%%%%%%%%%%%%%%%%%%%%%%%%%%%%%%%%%%
%%%%%%%%%%%%%%%%%%%%%%%%%%%%%%%%%%%%%%%%%%%%%%%%%%%%%%%

\section{Observations}\label{sec:obs}

2MASX J04035024-0239275 is a bright red elliptical galaxy at R.A.=04$^{\rm h}$03$^{\rm m}$50\fs26, Dec=$-$02$^\circ$39\arcmin27\farcs64 with a radial velocity $cz\sim$19,952 km s$^{-1}$ \citep{2009MNRAS.399..683J}, corresponding to an angular scale of 1.256 kpc arcsec$^{-1}$. 
The type Ia supernova LSQ13cwp exploded 9.6\arcsec~N and 6.1\arcsec~E from the galaxy core and was discovered on November 8th 2013 \citep{2013ATel.5567....1W}.
For this reason, 2MASX J04035024-0239275 was included in the sample of supernova host galaxies of the All-weather MUse Supernova Integral-field Nearby Galaxies (AMUSING; \citealt{2016MNRAS.455.4087G}) survey. 
AMUSING is an ongoing project aimed at studying the environments of supernovae (SNe) by means of the analysis of a large number of nearby SN host galaxies (redshifts typically between 0.005 and 0.1). 
The AMUSING sample currently comprises more than 300 galaxies observed in the last 6 semesters (P95 to P100) and is composed by a wide variety of galaxy types with the common characteristics of having hosted a known SN.

\begin{table}
\caption{Coordinates and redshifts of all galaxies found in the FoV, including the lens galaxy (A), the lensed source (B+C), and the merger (S).}
\label{tab:info}
\begin{center}
\begin{tabular}{clll}
\hline\hline
\textbf{Label} & \textbf{R.A.}  & \textbf{Dec}  & \textbf{$z$}  \\
\hline
A (lens)        & 4$^{\rm h}$03$^{\rm m}$50\fs24 & -02$^\circ$39\arcmin27\farcs4 & 0.0661 \\
B (Image I$_+$) & 4$^{\rm h}$03$^{\rm m}$50\fs16 & -02$^\circ$39\arcmin29\farcs2 & 0.1915 \\
C (Image I$_-$) & 4$^{\rm h}$03$^{\rm m}$50\fs25 & -02$^\circ$39\arcmin26\farcs6 & 0.1915 \\
D               & 4$^{\rm h}$03$^{\rm m}$50\fs79 & -02$^\circ$39\arcmin40\farcs2 & 0.4556  \\
E               & 4$^{\rm h}$03$^{\rm m}$50\fs53 & -02$^\circ$39\arcmin42\farcs0 & 0.4551  \\
F               & 4$^{\rm h}$03$^{\rm m}$49\fs71 & -02$^\circ$39\arcmin39\farcs6 & 0.4538  \\
G               & 4$^{\rm h}$03$^{\rm m}$49\fs49 & -02$^\circ$39\arcmin28\farcs2 & 0.4564  \\
H               & 4$^{\rm h}$03$^{\rm m}$48\fs74 & -02$^\circ$39\arcmin31\farcs2 & 0.4542  \\
I               & 4$^{\rm h}$03$^{\rm m}$49\fs30 & -02$^\circ$39\arcmin23\farcs8 & 0.4549  \\
J               & 4$^{\rm h}$03$^{\rm m}$49\fs79 & -02$^\circ$39\arcmin06\farcs3 & 0.4523  \\
K               & 4$^{\rm h}$03$^{\rm m}$49\fs59 & -02$^\circ$39\arcmin10\farcs6 & 0.0946  \\
L               & 4$^{\rm h}$03$^{\rm m}$49\fs77 & -02$^\circ$39\arcmin03\farcs8 & 0.0946  \\
M               & 4$^{\rm h}$03$^{\rm m}$52\fs02 & -02$^\circ$39\arcmin05\farcs5 & 0.5944  \\
N               & 4$^{\rm h}$03$^{\rm m}$50\fs69 & -02$^\circ$39\arcmin21\farcs0 & 0.5929  \\
O               & 4$^{\rm h}$03$^{\rm m}$50\fs27 & -02$^\circ$39\arcmin51\farcs6 & 0.5937  \\
P               & 4$^{\rm h}$03$^{\rm m}$50\fs51 & -02$^\circ$39\arcmin48\farcs4 & 0.6286  \\
Q               & 4$^{\rm h}$03$^{\rm m}$51\fs69 & -02$^\circ$39\arcmin06\farcs2 & 0.9031  \\
R               & 4$^{\rm h}$03$^{\rm m}$51\fs45 & -02$^\circ$39\arcmin07\farcs0 & unkn.   \\
S (merger)      & 4$^{\rm h}$03$^{\rm m}$50\fs67 & -02$^\circ$39\arcmin30\farcs0 & 0.0661  \\\hline
\end{tabular}
\end{center}
\end{table}

\begin{figure*}
\centering
\includegraphics[width=\textwidth]{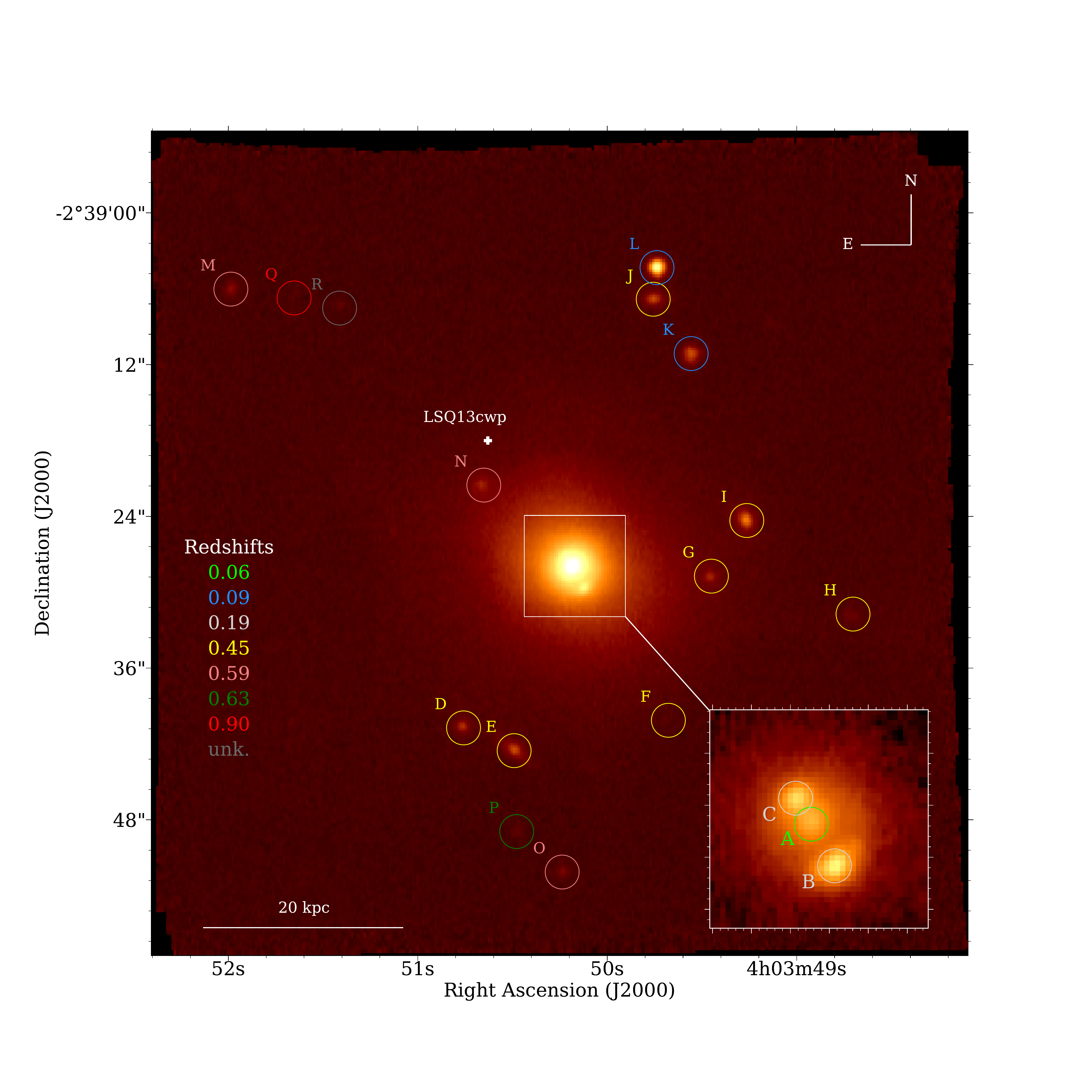}
\caption{Synthetic 10 \AA~narrow-band image (centered at 7184 \AA) of the MUSE observation. We identified 16 different sources in the field of view, indicated with circles. The inset is a synthetic narrow-band image centered at the H$\alpha$ emission of the source ($z$=0.1915) where the lens and the two images of the source can clearly be distinguished. Supernova LSQ13cwp position is represented by a white plus sign.}
\label{fig:fov}
\end{figure*}

2MASX J04035024-0239275 was observed using the integral-field spectrograph MUSE \citep{2010SPIE.7735E..08B} at the ESO Very Large Telescope (VLT) UT4 of the Cerro Paranal Observatory during a clear night on 2016 November 7th. 
In the Wide Field Mode, MUSE covers a continuous sky region of 1\arcmin$\times$1\arcmin~with a small spaxel size of 0.2\arcsec$\times$0.2\arcsec, which limits the spatial resolution of the data to the atmospheric seeing during the observation ($\sim$0.7\arcsec).
MUSE observations cover the wavelength range spanning from 4750\AA~to 9300\AA, with a spectral sampling of 1.25\AA~and a resolving power $R$ increasing from 1800 in the blue edge to 3600 in the red edge of the spectrum.
With its excellent total throughput, MUSE offers an unprecedented combination of sensitivity, spatial resolution, and  field of view (FoV).

\begin{figure*}
\centering
\includegraphics[width=\textwidth]{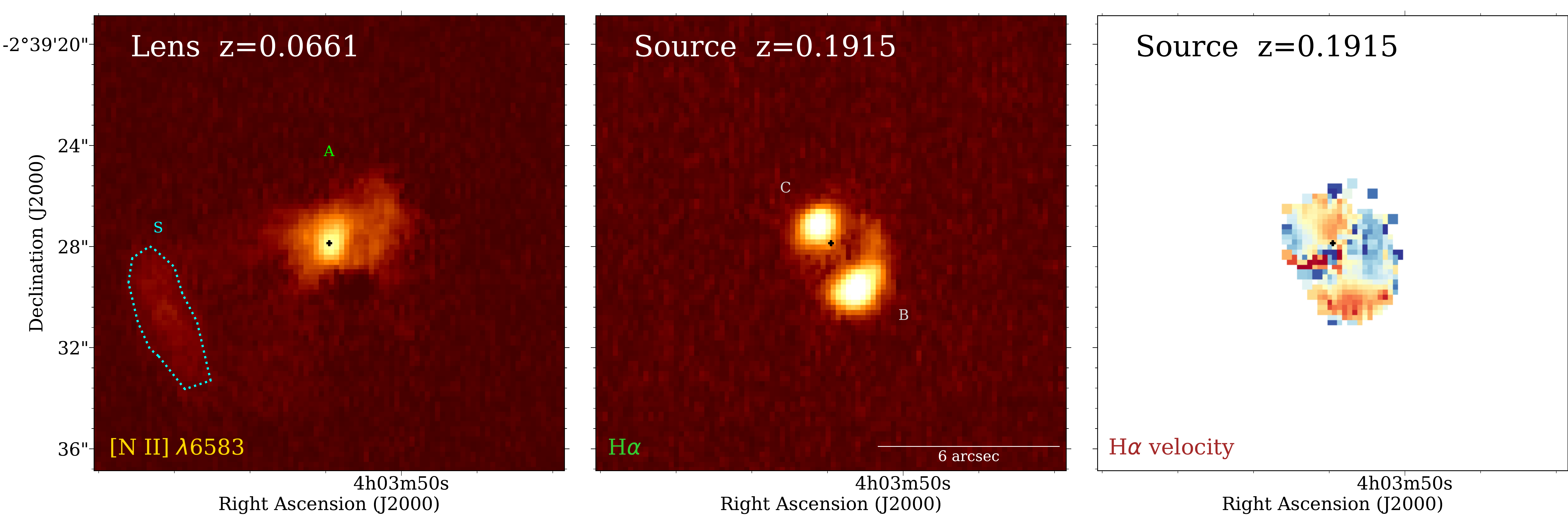}
\caption{22\arcsec$\times$22\arcsec~cutouts centered at the core of the foreground galaxy acting as a lens. 
$S$ region indicates the merger region (see section \ref{sec:merger}).
In the left panel, a [N II] $\lambda$6583 emission line flux map of the lens galaxy (7019 \AA~observer-frame), and in the center and right panels, H$\alpha$ flux and velocity maps of the background source galaxy (7820 \AA~observer-frame). An arc, forming a partial Einstein ring can clearly be identified in central and right panels.}
\label{fig:slices}
\end{figure*}

\begin{figure}
\centering
\includegraphics[trim=1.0cm 0cm 0.5cm 0cm, clip=true,width=0.92\columnwidth]{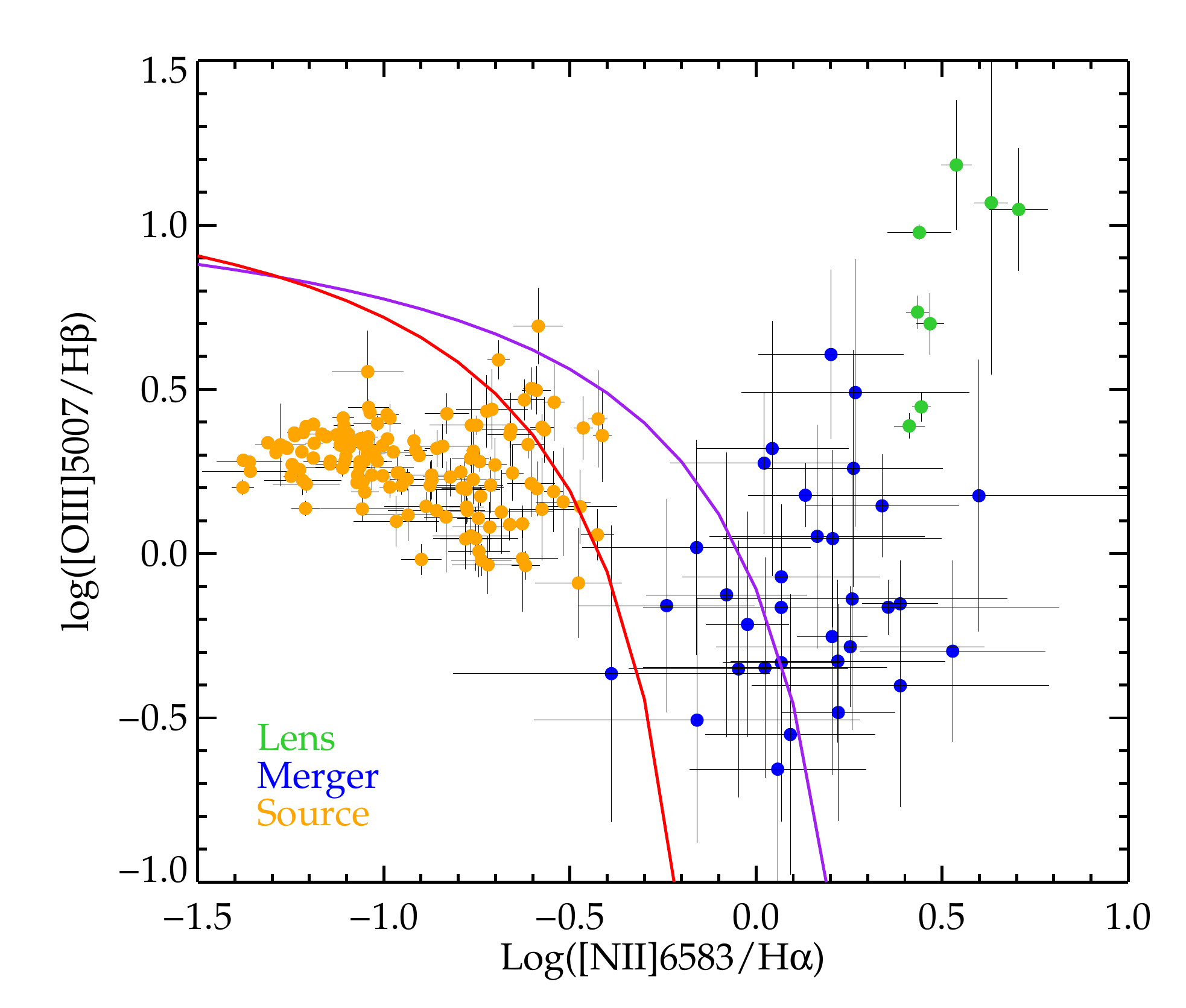}
\caption{BPT diagram constructed from line emission ratios measured in three distinct regions: the lens (A), the lensed source (B and C), and the possible merger (S). This diagram clearly shows that the origin of the ionization is different for the three objects. Purple and red solid lines correspond to the two criteria commonly used to separate star-forming (SF) from AGN-dominated galaxies from \protect\cite{2001ApJ...556..121K} and \protect\cite{2003MNRAS.346.1055K}, respectively.}
\label{fig:BPT}
\end{figure}

We obtained four dithered (by a few arcsec) exposures of 700s integration each, totaling 2800s on source, that resulted in a combined datacube with more than 105,000 individual spectra. 
There are no stars in the field of our MUSE observations, but we checked the image quality of the slow-guiding system (SGS) that uses metrology fields around the MUSE field of view (FoV) to provide a secondary telescope guiding. Three stars were detected in those fields with an average FWHM of 0.676 $\pm$ 0.040 arcsec.
We therefore assume throughtout the paper the point spread function (PSF) is a circular Gaussian with a full-width half-maximum of 0.7\arcsec, which corresponds to a physical resolution of 880 pc at the redshift of the lens, and it is consistent with the reported DIMM seeing. %This severely limits the fidelity of the lens modeling and our ability to robustly estimate statistical errors. 
%Given that there were no stars in the field to measure the full width half maximum of the point spread function (PSF), we used throughtout the paper the value reported by the DIMM during the observations, which was 0.7\arcsec and corresponds to a physical resolution of 880 pc at the redshift of the lens.
%
Data reduction was performed following prescriptions reported in \cite{2016MNRAS.455.4087G} and \cite{2017A&A...602A..85K}, and the reader is referred to those papers for details.

IFS can be thought as a sequence of two-dimensional (2D) very narrow-band imaging of the FoV covered by the instrument.
In the case of MUSE, this translates into 1.25\AA~wide 3,681 images.
Each of these images correspond to a different rest-frame wavelength for objects at different redshifts in the FoV.

Visual inspection of the  2MASX J04035024-0239275 datacube revealed a bright detached source located a few spaxels South-West of the core of the main galaxy.
A spectrum extracted with a small aperture at the position of the source resembled that of a typical star-forming galaxy at a higher redshift than the main galaxy ($z\sim$0.19).
Examining the wavelength slice (narrow-band image) of the datacube corresponding to the H$\alpha$ emission of this source (7820 \AA) we discovered another source at the North-Eastern side of the 2MASX J04035024-0239275 core (see inset of Fig. \ref{fig:fov}). 
The spectrum of this 2nd source was found to be very similar to the S-E object, and the estimated redshifts were consistent
Further inspection of the datacube revealed up to 15 other galaxies present in the FoV.
Figure \ref{fig:fov} shows a synthetic 10 \AA~narrow-band image from the datacube centered at 7184 \AA~in the observer frame\footnote{This wavelength was selected because it corresponds to the H$\alpha$ emission of galaxies $L$ and $K$. They only show a few bright emission lines on top of a very faint and noisy continuum (See Figure \ref{fig:BG1}), and otherwise they would not be visible in the image.}, where all objects detected are marked with circles of different colors grouped by their corresponding measured redshifts. 
Table \ref{tab:info} lists coordinates and spectroscopic redshifts of all objects detected in the FoV.
The inner panel presents a zoom-in of the 2MASX J04035024-0239275 core, from a synthetic narrow-band image centered at 7820 \AA, where it can be clearly seen the core (labeled as $A$ throughout the paper) and the two S-W (labeled as $B$) and N-E (labeled as $C$) bright blobs.

The geometrical configuration of $B$ and $C$ together with their similar spectra, suggested that they may be in fact the same object lensed by 2MASX J04035024-0239275. 
In the rest of the paper we present an analysis of the lens $A$ and the galaxy $B+C$, and demonstrate that this is actually the case.

%%%%%%%%%%%%%%%%%%%%%%%%%%%%%%%%%%%%%%%%%%%%%%%%%%%%%%%
%%%%%%%%%%%%%%%%%%%%%%%%%%%%%%%%%%%%%%%%%%%%%%%%%%%%%%%
%%%%%%%%%%%%%%%%%%%%%%%%%%%%%%%%%%%%%%%%%%%%%%%%%%%%%%%

\section{The lens}\label{sec:lens}

\begin{figure*}
\centering
\includegraphics[width=0.92\textwidth]{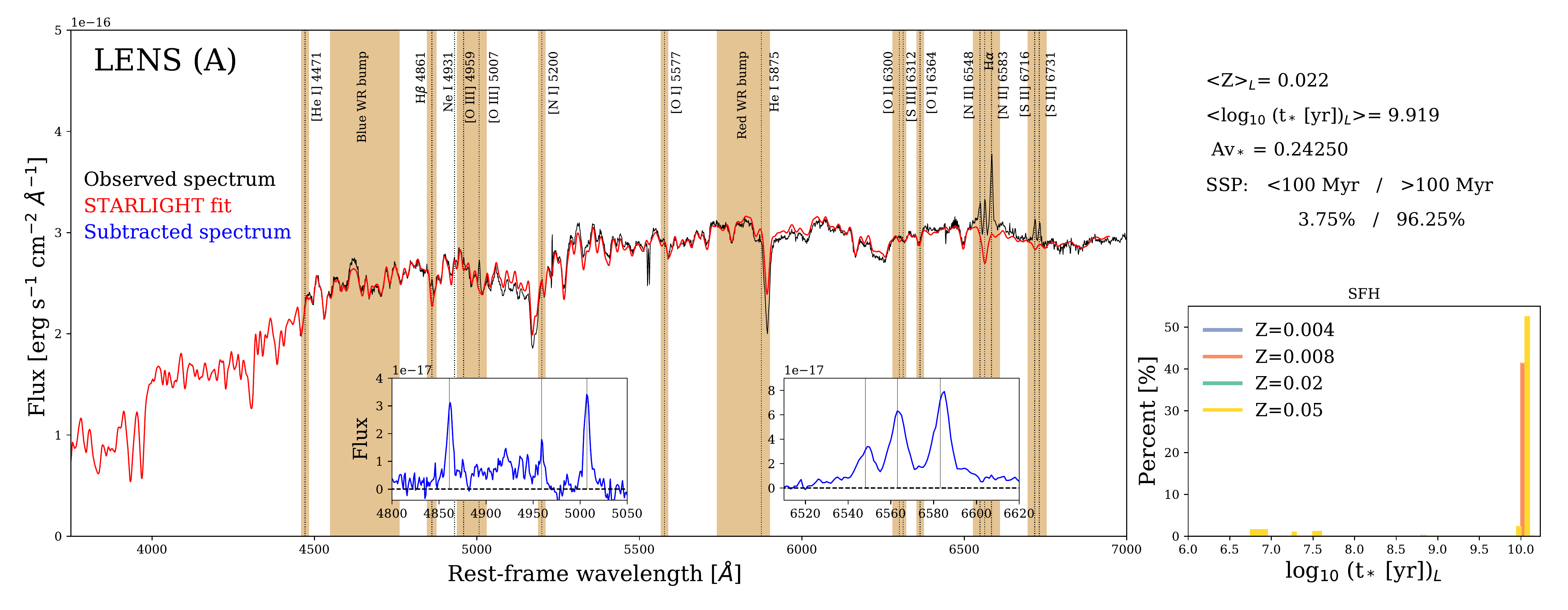}
\includegraphics[width=0.92\textwidth]{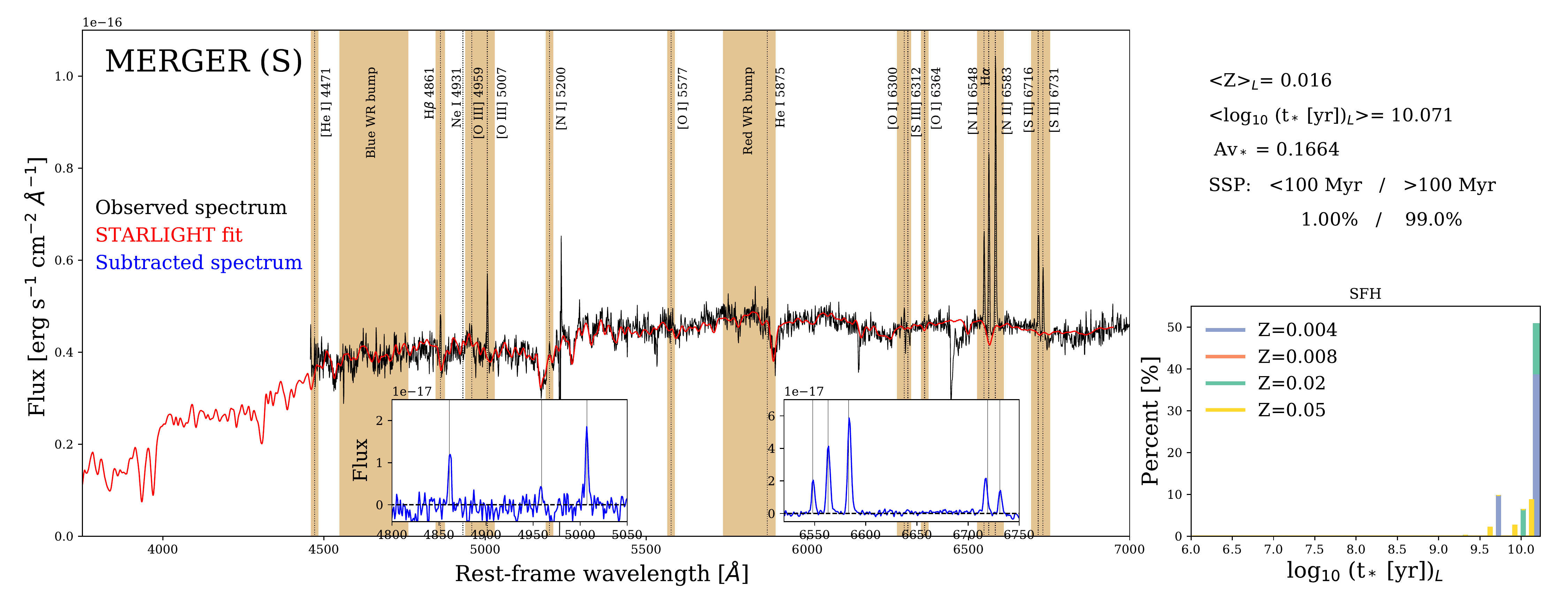}
\caption{Observed spectrum and STARLIGHT SSP best fit of the lens galaxy (top panel) and the possible merger (bottom panel).
Shaded areas are masked out in the fit as they correspond to the location of strong emission lines.
On the right side of each panel, the corresponding star formation history is shown with some summary parameters.}
\label{fig:lens_sp}
\end{figure*}

\subsection{Reconstructing the lens}

Although the images of the background galaxy are clearly seen in the observed cube, their visibility might be enhanced by subtracting a model of the contribution of the foreground galaxy acting as a gravitational lens. 
We have explored a number of approaches to model the spectrophotometric properties of the lens, namely: isophotal fitting, analytical S\'ersic fitting, and single stellar population (SSP) synthesis.

The main differences between these methods is the way they handle the stellar continua and the ionized gas emission.
While isophotal and S\'ersic fitting may contain traces of emission lines from the lens because gas and stars are treated together (at each wavelength slice), the SSP fitting model only contain the stellar component.
This contamination from the ionized gas is more important close to the lens core, and irrelevant at larger radii.
On the other hand, the intensity of the stellar continuum of the lens becomes fainter with radial distance, and SSP fitting does not provide reliable fits.
However, isophotal and S\'ersic fitting is noiseless and respects the shape of the lens stellar continuum.
Therefore, we use 
the {\it analytical S\'ersic fit} approach to remove the lens contribution from the spectra of all background galaxies (objects $D$ to $R$; See section \ref{app}), and
the SSP fitting approach to model the lens and the lensed system in the central region. % (See section \ref{sec:source}), and .

\begin{table*}
\caption{Extinction corrected flux of the most prominent gas phase emission lines and properties extracted from integrating the model datacube (for $A$), from aperture spectra in the residual cube after subtracting the lens model (for $B$ and $C$), and integrating the region defined by the dotted blue region area in Figure \ref{fig:slices} (for $S$). All fluxes in units of 10$^{-18}$ erg s$^{-1}$ cm$^{-2}$ $\AA^{-1}$. }
\label{tab:linesandprop}
\begin{center}
\begin{tabular}{lcccc}
\hline\hline
                                       &  A (lens)       &  B (Image I$_+$) & C (Image I$_-$) & S (merger) \\
\hline                
[O\,II]\,$\lambda\lambda$3727,29            &  $-$            & $-$              & $-$              & $-$             \\
H\,$\delta$                                 &  $-$            & 408.48 (28.14)  &  180.00 (27.59)  & $-$             \\
H\,$\gamma$                                 &  $-$            & 771.80 (60.97)  &  442.67 (39.02) & $-$             \\
H\,$\beta$                                  &  171.80 (26.90) & 1538.85 (47.76) & 879.77 (77.80)  & 83.05 (12.44)   \\
\protect [O\,III]\,$\lambda$4959                     &   78.19 (12.24) & 1019.50 (64.33) & 523.89 (43.61)  & 31.25 (7.46)    \\
\protect [O\,III]\,$\lambda$5007                     &  213.29 (33.39) & 2930.72 (92.57) & 1705.22 (44.40) & 95.85 (31.05)   \\
\protect [N\,II]\,$\lambda$6548                      &  159.63 (24.99) & 467.93 (43.41)  & 284.88 (25.78)   & 116.74 (34.98)  \\
H\,$\alpha$                                 &  369.65 (57.88) & 4401.12 (121.20)& 2516.15 (71.69) & 237.52 (57.22)  \\
\protect [N\,II] \,$\lambda$6583                     &  452.30 (70.82) & 1761.82 (66.67)   & 1025.20 (34.71)  & 353.33 (82.88)  \\
\protect [S\,II]\,$\lambda$6717                      &  182.16 (28.52) & 781.52 (97.46)  & 472.91 (50.56)  & 117.79 (29.73)  \\
\protect [S\,II]\,$\lambda$6731                      &  116.31 (18.21) & 713.79 (80.35)  & 445.37 (53.56)  & 81.26 (20.51)   \\
\hline                    
A$_V^{\rm gas}$ [mag]                       &  0.00 (0.02)    & 1.07 (0.03)      & 1.17 (0.04)      & 0.47 (0.03)     \\
12+log$_{10}$ (O/H) [dex]                   &  $-$            & 8.39 (0.09)      &  8.39 (0.10)     & $-$             \\
log$_{10}$ (SFR [M$_\odot$ yr$^{-1}$]) [dex]&  $-$            & 4.261143         & 2.436128         & $-$             \\
log$_{10}$ (M [M$_\odot$]) [dex]            &  11.54          & 10.35            & 10.13            & 9.79            \\
A$_V^*$ [mag]                               &  0.24           & 0.27             & 0.02             & 0.17            \\
<log$_{10}$ (t$_*$ [yr])> [dex]             &  9.92 (0.74)    & 7.31 (0.74)      & 7.04 (0.66)      & 10.07 (0.18)    \\
<log$_{10}$ Z$_*$> [dex]                    &  0.022 (0.016)  & 0.017 (0.013)    & 0.012 (0.014)    & 0.016 (0.014)   \\
\hline
\end{tabular}
\end{center}
\end{table*}

Ignoring the signal of all background sources, including the lensed source, we analyzed the lens following the procedures of \cite{2014A&A...572A..38G,2016A&A...591A..48G,2018ApJ...855..107G}. 
After correcting all spectra in the datacube for Galactic extinction ($E(B - V)=0.134$ mag) using dust maps from \cite{2011ApJ...737..103S} and assuming a \cite{1989ApJ...345..245C} extinction law with R$_V$=3.1, 
we fit all spectra in the observed datacube with STARLIGHT \citep{2005MNRAS.358..363C} fixing the redshift at 0.0661, corresponding to the recession velocity of the lens.
{\sc STARLIGHT} estimates the fractional contribution of different SSPs with different ages and metallicities to the stellar continuum in the spectra, assuming that the star formation history of a galaxy can be approximated as the sum of discrete star formation bursts.
The fitting makes use of the ``Granada-Miles'' (GM) base, which is a combination of the MILES SSP spectra provided by \citet[][as updated by \citealp{2011A&A...532A..95F}]{2010MNRAS.404.1639V} for populations older than t = 63 Myr and the \cite{2005MNRAS.357..945G} models for younger ages. They are based on the \cite{1955ApJ...121..161S} Initial Mass Function and the evolutionary tracks of \cite{2000A&AS..141..371G}, except for the youngest ages (<3 Myr), which are based on Geneva tracks \citep{1992A&AS...96..269S,1993A&AS...98..523S,1993A&AS..102..339S,1993A&AS..101..415C}. The GM base is defined as a regular (t, Z) grid of 248 models with 62 ages spanning t = 0.001-14 Gyr and four metallicites (Z/Z$_\odot$ = 0.2, 0.4, 1, and 1.5).

Given the relatively faint stellar and ionized emission of all background galaxies with respect to the lens, emission lines from background galaxies do not introduce significant contamination in the SSP fitting of the lens.
All individual STARLIGHT best fits are then put together in a 3D datacube that describes a model of the lens galaxy. 

\subsection{Separating the background source from the lens}

In order to separately study the lens and the background source, we removed the contribution of the lens from the source by subtracting the SSP fitting datacube from the 3D observation.
The residual contains the ionized emission lines from the lens (consistent with a redshift 0.0661) on top of both the stellar and the gas-phase emission of the source (at redshift 0.1915).
We fit all strong emission lines of ionized gas from the lens and the background source (H$\beta$, [O III] $\lambda$5007, [N II] $\lambda$6548, H$\alpha$, [N II] $\lambda$6583) using {\sc Pipe3D} \citep{2016RMxAA..52...21S}. %The resulting fluxes are reported in Table \ref{tab:linesandprop}.

For the lens, the [N II] $\lambda$6583 emission is stronger than H$\alpha$ suggesting that the underlying ionization is caused by an AGN. 
In Figure \ref{fig:slices} (left panel) we show the [NII] $\lambda$6583 2D emission line map of the lens, where the two blobs from the background galaxy are not present.
The AGN origin is confirmed by the location of the emission line ratios in the AGN region on the BPT diagram shown in Figure \ref{fig:BPT}.

On the other hand, the lensed source object shows strong H$\alpha$ emission indicating recent star-formation.
In the central panel of Figure \ref{fig:slices} we show the H$\alpha$ 2D emission line map, where the two images and a ring structure can clearly be seen. 
We note that the region within the images of the source appear clean of emission from the lens, confirming that the SSP approach works better in the lens core region.
Similarly, the origin of the ionization as star-forming activity is confirmed by the location of the flux ratios in the BPT diagram (See Figure \ref{fig:BPT}).

By determining the central wavelength of the H$\alpha$ emission of the source in each individual spaxel, we can reconstruct the velocity field of the two images.
In the right panel of Figure \ref{fig:slices} we show how the projected rotational velocity along the line-of-sight is in the opposite direction in each of the images.
We found velocities in the range $v \in [ -50, 50]$ km s$^{-1}$.

\begin{figure*}
\centering
\includegraphics[trim=0cm 0cm 0cm 0cm,clip=True,width=\textwidth]{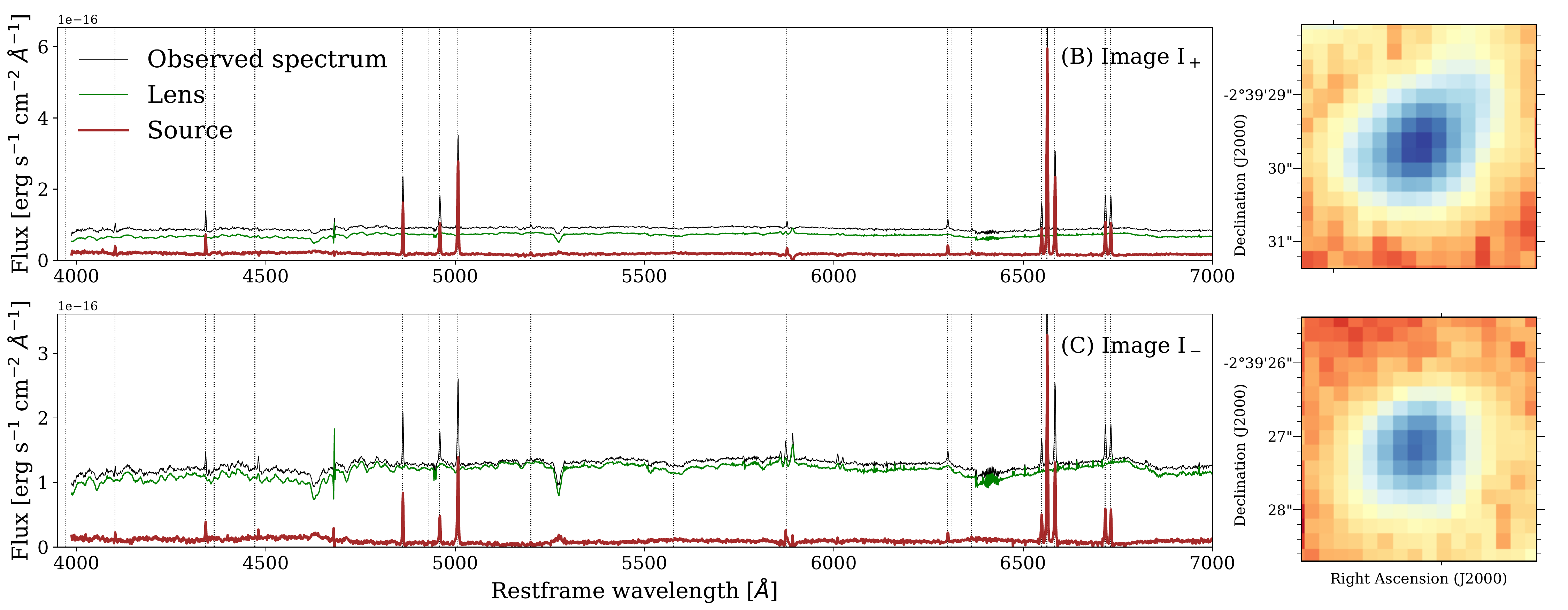}
\caption{Spectra of images I$_+$ (source $B$) and I$_-$ (source $C$). In black we show the observed spectrum obtained from a 2 arcsec diameter aperture, in green the model of the lens extracted from an aperture of the same size in the synthetic model datacube, and in red the residual. On the right side, the H$\alpha$ emission 2D map of both sources.}
\label{fig:BC}
\end{figure*}

\subsection{Possible merger remnant}
\label{sec:merger}

Another interesting result from our SSP fitting that is also evident in the left panel of Figure \ref{fig:slices}, is the presence of a residual shell-like structure on the eastern side of the lens (labeled as $S$ hereafter). From inspection of the original cube, this structure is clearly at the same redshift of the lens, although it presents emission lines from ionized gas. The integrated spectra of the lens model and that corresponding to the source encircled by the dotted blue region area represented in Figure \ref{fig:slices}, are shown in black in Figure \ref{fig:lens_sp}. They both have red stellar continua characteristic of old stellar populations with faint gas emission lines lying on top. The stellar continuum is noisier than the lens core but we still get a meaningful SSP fit from STARLIGHT that is shown in red. The two inner panels  zoom-in to the wavelengths around H$\beta$ and H$\alpha$. In blue we show the difference between the observation and the SSP fit, where emission lines can clearly be identified.
On the right side, the reconstructed star formation history is presented, which shows that 96.25\% and 99\% of the SSP models needed for the best fit of the lens and the merger, respectively, correspond to populations of ages older than 3 Gyr ($\sim$9.5 dex).
Stellar parameters from the SSP synthesis such as the average stellar age, metallicity, stellar extinction and stellar mass, as well as properties estimated from the gas such as the gas extinction, oxygen abundance and on-going star formation rate are reported in Table \ref{tab:linesandprop}.

The presence of ionized gas in early-type galaxies is known to be more frequent than what was considered in recent times \citep[e.g.][]{sarzi10,papa13,sign13}. The source of the ionization is generally a mix between different physical processes, including four major sources: (i) nuclear activity (i.e., AGN), that have a characteristic ionization with strong [N II]/H$\alpha$ and [O III]/H$\beta$ ratios \citep[e.g.][]{Gomes16a}; (ii) outflows by sometimes undiscovered central AGNs \citep[e.g.][]{kehrig12,Cheung16}; (iii) ionization by young stars, remnants of a dimming disk or a consequence of a rejuvenation by the capture of a gas rich satellite galaxy \citep[e.g.][]{Gomes16b}; and, in most of the cases, (iv) diffuse ionization by post-AGB stars \citep[][]{binn94,sta08,sarzi10,papa13,sign13,Gomes16a,belf16a}. Although the nature of the ionization is more or less clear, it is still unclear the origin of this ionized gas. Although there are elliptical galaxies without trace of either ionized or cold gas, many of them show both phases of the gas \citep[][]{sanchez17b}. Thus, maybe this gas represents the left-overs of the early evolution of these galaxies, that once were gas rich, star-forming, and have halted their activity due to gas starvation \citep[e.g.][]{bekki02}.

However, the detected structure, named as $S$, comprises both ionized gas and stellar populations, and it presents a shell-like structure located at the SE of the lens, with a diffuse (smoother) distribution towards the center of the galaxy. Its morphology is different from the one expected either by a direct ionization by an AGN or a centrally distributed gas, or the spiral-like structure of ionization due to young stars. Its location in the BPT diagram indicates that the source of the ionization is LINER-like, either associated with post-AGBs or shocks (See Figure \ref{fig:BPT}). Even more, shell-like structures are characteristic of past merging events, that disrupted the morphology of the captured galaxy and/or created resonances in the distorted disk \citep[e.g.][]{bego05}. Altogether we consider that this structure is totally unrelated to the lensed galaxy, being the remnant of a past merging event. We consider that the lens galaxy has captured, disrupted and integrated a less massive object, more gas-rich, and it has produced the observed structure due to dynamical processes. The gas remnant is ionized either by post-AGB stars, shocks or a combination of both processes. 
%This structure, while interesting by itself, does not provide further information on the nature of the lensed galaxy.

%%%%%%%%%%%%%%%%%%%%%%%%%%%%%%%%%%%%%%%%%%%%%%%%%%%%%%%
%%%%%%%%%%%%%%%%%%%%%%%%%%%%%%%%%%%%%%%%%%%%%%%%%%%%%%%
%%%%%%%%%%%%%%%%%%%%%%%%%%%%%%%%%%%%%%%%%%%%%%%%%%%%%%%

\begin{figure*}
\centering
\includegraphics[trim=0cm 0cm 0cm 0cm,clip=True,width=\textwidth]{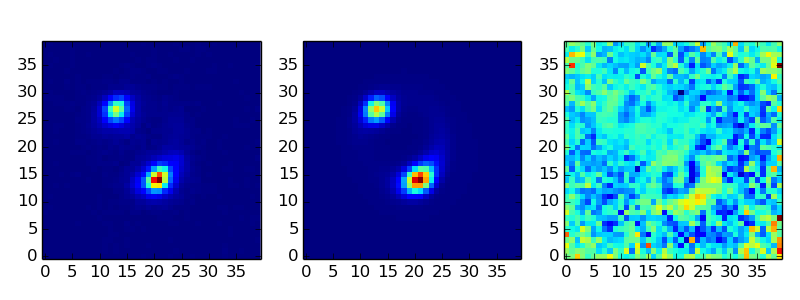}
\caption{Best fit reconstruction of the lensed images assuming a SIE lens model and a single S\'ersic profile for the source. Coordinates are in 0.2" MUSE pixels. From left to right: data, model, residual. The residuals are poor because the PSF is not well known - there are no stars in the field of view.}
\label{fig:reconstruction}
\end{figure*}

\section{The background source} \label{sec:source}

We estimated the position of the galaxy lens core and the center of the two source images by fitting a 2D gaussian profile in the [N II] $\lambda$6583 and the H$\alpha$ emission map, respectively (See Figure \ref{fig:slices}).
We obtained an angular distance from the lens core to the south-western (B) image of $\sigma_+$=2.056 arcsec (1.58 kpc), and to the north-eastern image of $\sigma_-$=$-$1.013 arcsec (1.27 kpc). 

To study the source we extracted 2\arcsec-diameter aperture spectra at the position of the two $B$ and $C$ images, in both the observed datacube and in the synthetic lens model.
In Figure \ref{fig:BC} we show the observed spectra of the $B$ and $C$ images in black lines, the lens spectrum in green, and the subtraction residual in red. On the right, the H$\alpha$ emission 2-dimensional map of both sources is shown.
STARLIGHT fit results of the residual spectra are listed in Table \ref{tab:linesandprop}.

Light from both images suffer different amounts of extinction while traveling through different sides of the lens. 
A larger amount of extinction is reflected in a higher observed $H\alpha$/$H\beta$ ratio, which once corrected increases the flux of H$\beta$ and [O III] $\lambda$5007 over H$\alpha$ and [N II] $\lambda$6583, but does not affect (significantly) both [O III]/H$\beta$ and [N II]/H$\alpha$ ratios.
$A_V$ measured from the Balmer decrement, H$\alpha$ over H$\beta$ observed ratio, assuming an $R_V$ of 3.1, is 1.07 (0.03) mag for $B$ and 1.17 (0.04) mag for $C$. 
In addition, the stellar extinction obtained from the SSP fitting $A_V^*$ is 0.27 for $B$ and 0.02 for $C$.

Oxygen abundance is measured in the two spectra using the O3N2 calibrator from \cite{2013A&A...559A.114M}.
Interestingly, we find the same value from the spectra of two images, 8.39 dex.
This is expected and confirms that the two images correspond indeed to the same source galaxy.
On the other hand, the star formation rate (SFR) estimated through the expression provided by \cite{1998ARA&A..36..189K} using the H$\alpha$ emission flux, is a factor 1.75 higher for $B$ than for $C$.
We argue below that this different strength of the emission is due to the magnification effect, which is in turn different for each image.

\section{Lens Modeling}
\label{sec:lensmodelling}

\begin{figure}
\centering
\includegraphics[trim=1cm 1cm 1cm 1cm,clip=True,width=\columnwidth]{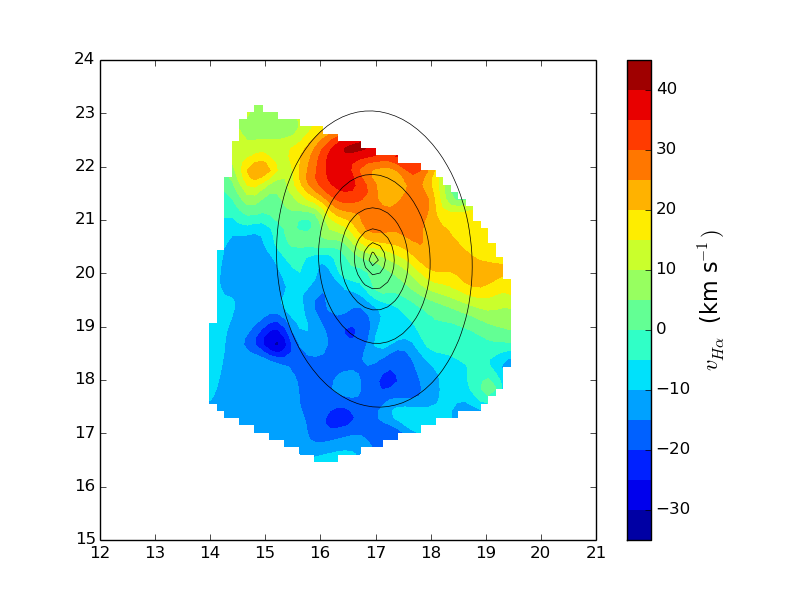}
\caption{Velocity reconstruction of the source from tracing the observed H$\alpha$ velocity onto the source plane using the best fit lens model. The best fit reconstructed light profile is indicated by the black contours.}
\label{fig:reconstruction2}
\end{figure}

The observed positions and fluxes of the multiple images are sensitive to the total lensing mass and its distribution within the lens. Since the observed images are extended, families of lens models that reproduce the observed image separations, but not the shape of the lensed features can be excluded.

Extracting this information requires a detailed reconstruction of the observed source light profile. By ray-tracing all of the observed image plane pixels through a lensing mass model back on the source plane to reconstruct a single self-consistent source.

To reconstruct the observed light profile of the arcs, we fit a singular isothermal ellipsoid (SIE) for the lensing mass and a single elliptical S\'ersic profile for the background source. We allow the Einstein radii, positions, position angle, and flattening of the lens to be free parameters and we do not fix any to the observed lens positions.

\begin{figure*}
\centering
\includegraphics[trim=0cm 0cm 0cm 0cm,clip=True,width=\textwidth]{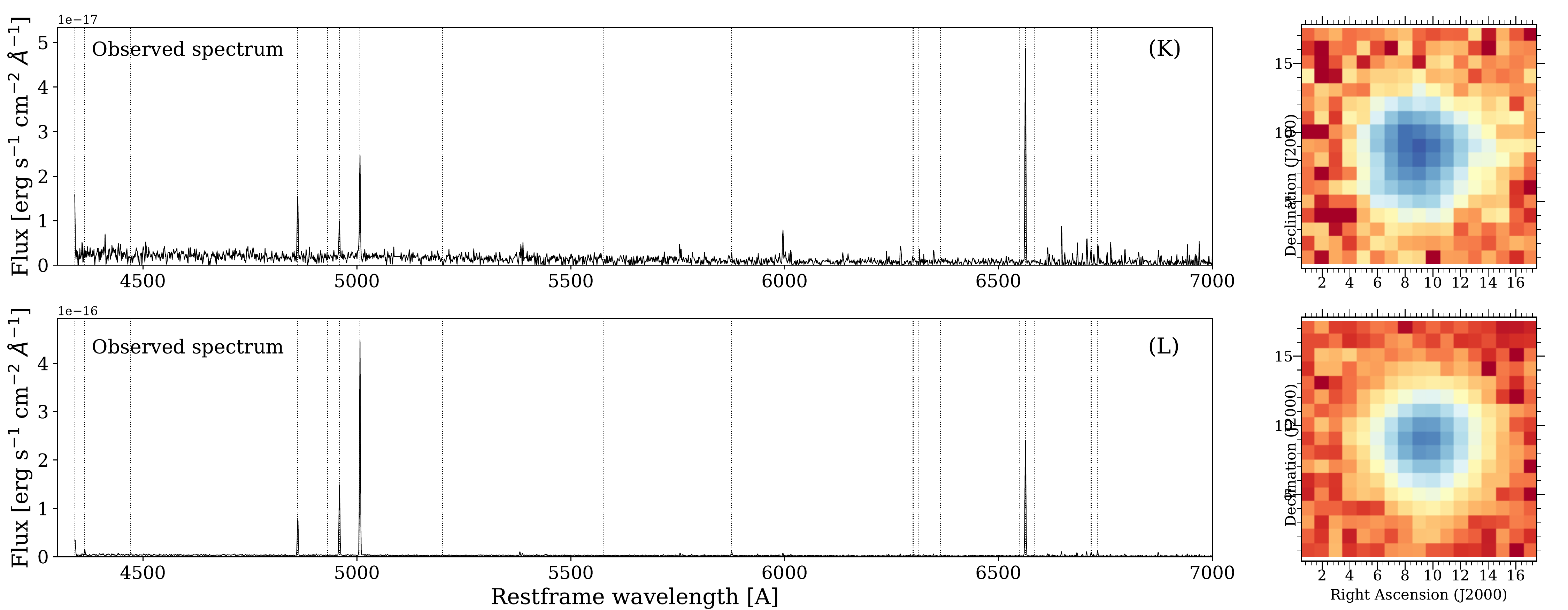}
\caption{Spectra of background galaxies $K$ and $L$ both at $z$=0.09. They show very faint stellar continua and 4 prominent emission lines. On the right, we show cutouts of the H$\alpha$ 2D emission line map.}
\label{fig:BG1}
\end{figure*}

The model is able to reconstruct the observed light profile, including the tail of the arc to the North West of image B. The best fit model is shown in Figure \ref{fig:reconstruction}. %The significant residuals in this figure are due to the PSF mismatch.  
The Einstein radius is 1.45 $\pm$ 0.04 arcsec. 
We estimate errors from fitting with a PSF of 0.6\arcsec~and 0.8\arcsec~FWHM. The inferred lens center of mass and the flattening and ellipticity are consistent with the observed light profile. %, but the ellipticity has significant uncertainty due to lack of robust PSF.

The observed Einstein radius corresponds to a physical scale of 1.9 kpc and an enclosed lensing mass of $1.08 \pm 0.06 \times 10^{11} M_\odot$. 

\subsection{Dark matter fraction}

Comparing the lensing and the stellar masses allows us to estimate the dark matter mass enclosed within the Einstein radius.
To measure the dark matter fraction, we extracted a circular aperture spectrum from the 3D datacube of the STARLIGHT model of the lens (second approach described in section \ref{sec:lens}) with an angular size of the Einstein radius (1.45\arcsec), and analyzed the spectrum with STARLIGHT to estimate the stellar mass.

The best SSP synthesis provides a star formation history with an average stellar age of 2.6 Gyr and an average metallicity of 1.1 Z$_\odot$, with two dominant bursts at log$_{10}$ (t [yr]) $\sim$9.4 and 10.1 dex. We obtained log$_{10}$ M$^*_{<\theta_E}$ [M$_\odot$] = 10.946 $\pm$ 0.131 dex, which corresponds to 82\% $\pm$ 8 \% of the mass enclosed within the Einstein radius, and means that dark matter contributes 18\% $\pm$ 8 \% to the total mass within the Einstein ring. 
This fraction is consistent with other reports in the literature for galaxies with similar stellar mass \citep{2013MNRAS.432.1709C}.

%\subsection{IMF mass excess parameter}
%
% convolved the spectrum to SDSS/PaNSTARRS i-band
% we find 15.71 mag -> L=2.9051558e+10 Lsolar ->  M/L of 3.03 in solar units
% M/L_ref below 2 ->  M/L / M/L_ref ~ 1.5

\subsection{Velocity field and dispersion }

By ray-tracing through the best fit lens model, we can reconstruct the velocity field of the source. This reconstruction is shown in Figure \ref{fig:reconstruction2}: The source shows a velocity gradient across the source, with the rotation axis offset from the inferred semi-major axis of the source. A more detailed reconstruction of both lens and source is required to investigate if this offset is real. 

Assuming an isothermal lens, the Einstein radius predicts a velocity dispersion of $280 \pm 4$ km~s$^{-1}$.
Our STARLIGHT fit for the lens model within the Einstein radius has a measured velocity dispersion of $\sigma_v^{\rm STAR}$ = 247 $\pm$ 32 km s$^{-1}$, once corrected for instrumental effects and the difference in dispersion of the bases used for the SSP synthesis. 
The corresponding dispersion measured in the observed spectrum within the Einstein radius is larger  298 $\pm$ 29 km s$^{-1}$, although in this latter case the spectrum includes emission lines and contamination from the background source that affects the SSP fitting.
These results are indicative that the total density profile is approximately isothermal \citep{2010ApJ...724..511A}.

%%%%%%%%%%%%%%%%%%%%%%%%%%%%%%%%%%%%%%%%%%%%%%%%%%%%%%%
%%%%%%%%%%%%%%%%%%%%%%%%%%%%%%%%%%%%%%%%%%%%%%%%%%%%%%%
%%%%%%%%%%%%%%%%%%%%%%%%%%%%%%%%%%%%%%%%%%%%%%%%%%%%%%%

\begin{figure*}
\centering
\includegraphics[trim=0cm 0cm 0cm 0cm,clip=True,width=0.85\textwidth]{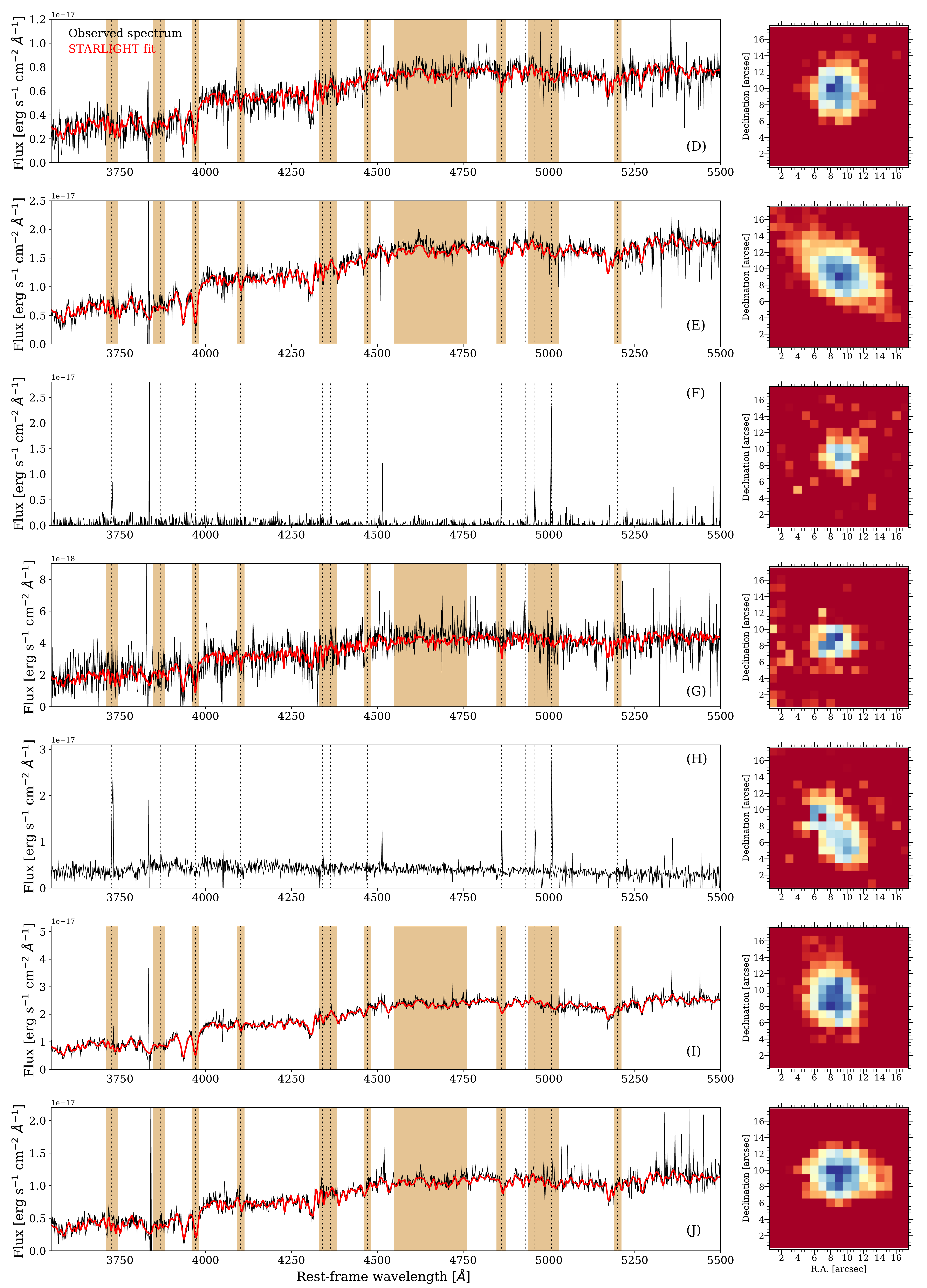}
\caption{Spectra of background galaxies $D$ to $J$, all at redshifts $z\sim$0.45. We show the observed spectrum in black and the best STARLIGHT fit in red. Brown regions are masked in the STARLIGHT fit, while the wavelengths of the most prominent emission lines are marked with vertical dotted lines. On the right, we show 4\arcsec$\times$4\arcsec~cutouts of the S/N 2D map. For galaxies $F$ and $H$ we show the [O III] 
$\lambda$5007 emission map instead.}
\label{fig:BG2}
\end{figure*}

\begin{figure*}
\centering
\includegraphics[trim=0cm 0cm 0cm 0cm,clip=True,width=0.9\textwidth]{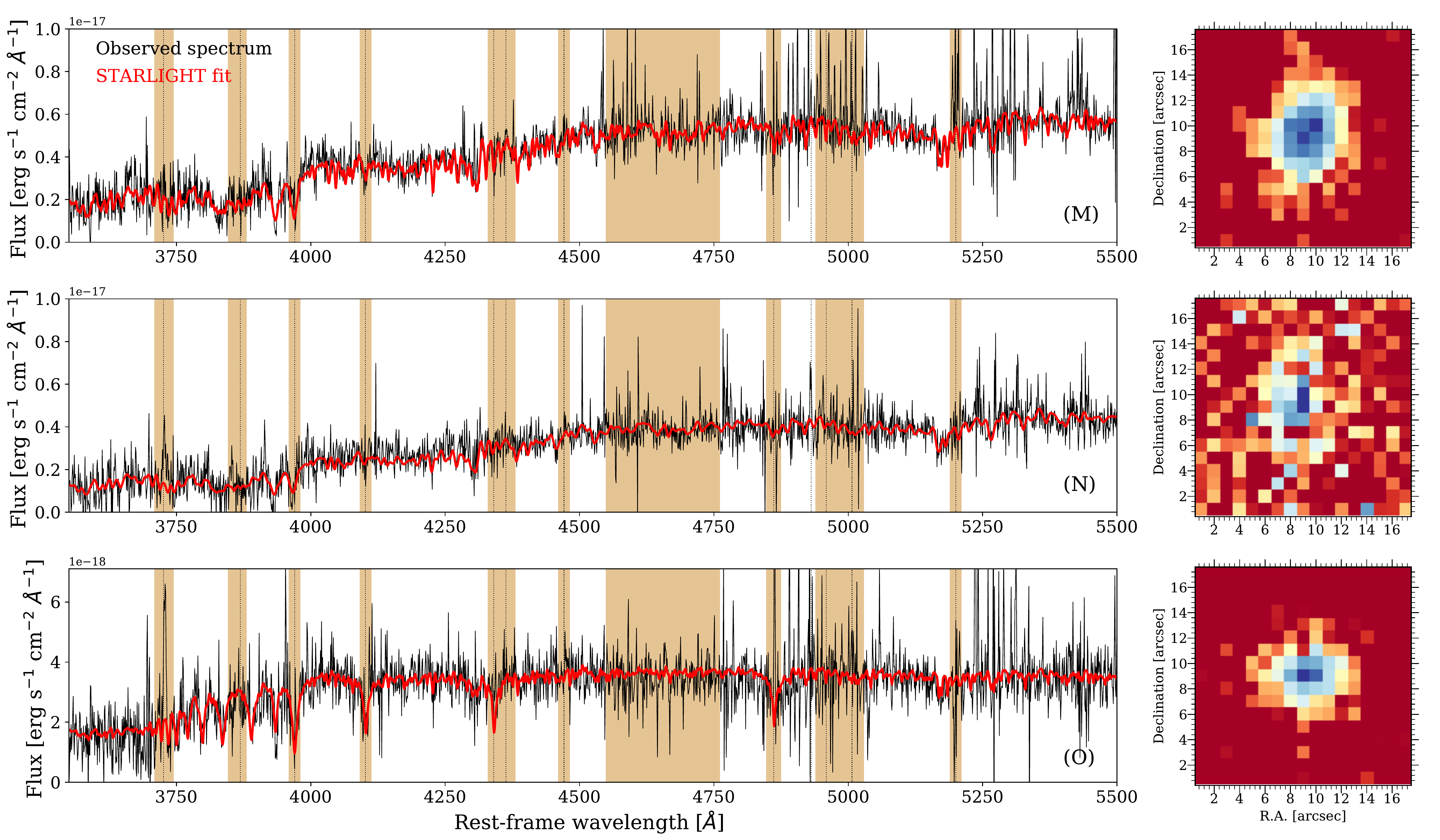}
\caption{Similar to Figure \ref {fig:BG2} for galaxies $M$, $N$, and $O$ detected at redshifts $z\sim$0.59.}
\label{fig:BG3}
\end{figure*}

\begin{figure*}
\centering
\includegraphics[trim=0cm 0cm 0cm 0cm,clip=True,width=0.9\textwidth]{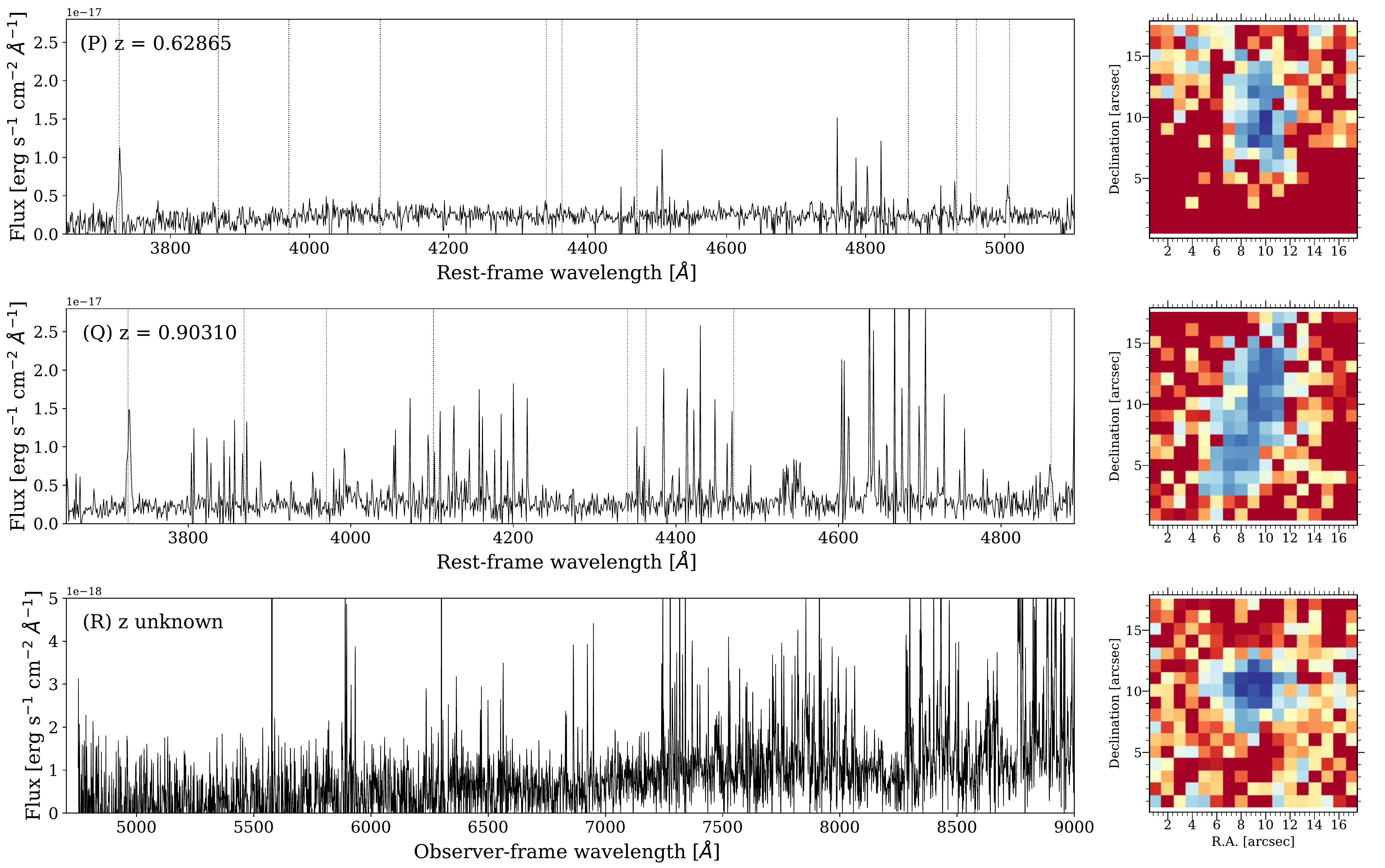}
\caption{Similar to Figure \ref {fig:BG2} for galaxies $P$ and $Q$ detected at redshifts $z\sim$0.62 and 0.90, respectively, and object $R$ at unknown redshift.}
\label{fig:BG4}
\end{figure*}

\section{Other galaxies in the MUSE field} \label{app}

We have used an analytical S\'ersic fitting approach to remove the lens contribution from the spectra of all other galaxies in the field of view.

The way to model the spectrophotometric properties of the lens is using the two-dimensional photometric decomposition code GASP2D \citep{mendezabreu08,mendezabreu14}. It allows to fit a S\'ersic model to the galaxy surface brightness distribution, thus deriving the main structural parameters of the galaxy, i.e., effective surface brightness ($\mu_e$), effective radius ($r_{\rm eff}$), S\'ersic index ($n$), ellipticity ($\epsilon$), and position angle ($PA$). GASP2D uses an iterative Levenberg-Marquardt algorithm to find the solution that minimise the $\chi^2$. At each iteration, the galaxy model is convolved with the image Point Spread Function (PSF) before computing the $\chi^2$. In our case we consider the PSF to be a circular Moffat with Full Width at Half Maximum FWHM=0.7 arcsec and shape parameter $\beta=1.95$. Similarly as with the isophotal fitting, each wavelength slice of the datacube is treated as an independent galaxy image, and the best fit S\'ersic parameters are derived. The averaged S\'ersic parameters over the whole wavelength range of the model are: $r_eff$= 8.8\arcsec (9.30 kpc), $n$=5.2 ,$\epsilon$=0.1, and $PA$=51 deg (N$\rightarrow$E). It is worth noting that a test with a S\'ersic+Exponential model was also performed. However, it did not improve the single S\'ersic fit and we use the latter to avoid over-fitting.

Once the model is subtracted from the observed datacube, we have found 15 bright sources within the FoV, besides the lens, the source, and the merger.
Table \ref{tab:info} lists their coordinates and spectroscopic redshifts of all objects detected in the FoV.
To investigate the nature of their emission, we have extracted 1\arcsec~diameter circular aperture spectra at those positions in both the observed and the lens model datacubes and studied the residual spectra.
We have been able to characterize 14 of these 15 sources, and details are given below.

Two sources, labeled $K$ and $L$ in Figure \ref{fig:fov}, show only 4 strong ionized gas emission lines corresponding to star-forming galaxies at redshift 0.09, as shown in Figure \ref{fig:BG1}.
In this Figure we also present 4\arcsec$\times$4\arcsec~cutouts of the H$\alpha$ 2D emission line map.
H$\alpha$ is the strongest line in the spectrum of $K$, which also shows H$\beta$, [OIII] $\lambda$4959, and [OIII] $\lambda$5007. 
Galaxy $L$ has the same 4 lines but in this case [OIII] $\lambda$5007 is stronger than H$\alpha$.
In the blue end of the spectrum there are indications of H$\gamma$ in both spectra, but only the red wing is visible.
[N II] $\lambda\lambda$6548,6583 doublet is not present in any of the two galaxies, which avoids estimating oxygen abundance through typical empirical calibrations.
However, in both spectra [OIII] $\lambda$5007 is stronger than H$\beta$ indicating low-metallicity.
No STARLIGHT fit was performed due to the faint stellar continuum, but fluxes of all prominent emission lines are reported in Table \ref{tab:otherelines}.

Further than $K$ and $L$, we find the lensed source galaxy at z=0.1915, and following that we detected 7 different sources at redshifts around 0.45.
%\textbf{[HK: I wonder if these galaxies at the same z can be considered as a cluster? what is the linear area they cover?]}
%
They are all marked in Figure \ref{fig:fov} with yellow circles, and labeled with letters $D$ to $J$.
In Figure \ref{fig:BG2} we present all their spectra extracted from a 1~arcsec diameter aperture, together with 4\arcsec$\times$4\arcsec~cutouts of the S/N of the spectra as measured in the observed-frame 50 \AA~window centered at 6000 \AA.
Five objects had a bright red continuum and no emission lines representative of passive galaxies dominated by old stellar populations. We fit these spectra with STARLIGHT and provide their properties in Table \ref{tab:otherprop}.
Objects $F$ and $H$ have significantly fainter continua and present a few strong emission lines.
For this reason, in Figure \ref{fig:BG2} we show the 2D [O III] $\lambda$5007 map instead of a S/N 4\arcsec$\times$4\arcsec~cutout.
In both galaxies, the spectrum does not cover the wavelength region corresponding to H$\alpha$ in the rest-frame, and [O III] $\lambda$5007 is the strongest emission line. Again, this line is stronger than H$\beta$ indicating metal-poor nature.

The next set of galaxies, $M$, $N$, and $O$ in Figure \ref{fig:fov}, are at redshift $\sim$0.59.
All their spectra, shown in Figure \ref{fig:BG3}, are compatible to passive red galaxies at $z\sim$0.59. 
We were able to fit STARLIGHT to the three spectra, and their properties are listed in Table \ref{tab:otherprop}.

Finally, in Figure \ref{fig:BG4} we present the spectra and cutouts of the other 3 galaxies found in the FoV.
The S/N of these spectra are too low for fitting SSP with STARLIGHT, but we are able to determine the redshift from their emission lines in two cases: 
object $P$ is at $z$=0.6286 according to the [OII] $\lambda\lambda$3727,29 and [OIII] $\lambda$5007 lines, and object $Q$ is at 0.9031 from [OII] $\lambda\lambda$3727,29 and H$\beta$ lines.
We have not been able to find good matches for object $R$, although we are convinced of the detection from its red continuum and the blob in the S/N map.
New, deeper spectroscopic observation is needed to provide further information for object $R$.

%%%%%%%%%%%%%%%%%%%%%%%%%%%%%%%%%%%%%%%%%%%%%%%%%%%%%%%
%%%%%%%%%%%%%%%%%%%%%%%%%%%%%%%%%%%%%%%%%%%%%%%%%%%%%%%
%%%%%%%%%%%%%%%%%%%%%%%%%%%%%%%%%%%%%%%%%%%%%%%%%%%%%%%

\section{Summary and discussion}\label{sec:conc}

We reported the discovery of a strong lensed galaxy at $z$=0.1915 by the lens 2MASX J04035024-0239275 at $z$=0.0661 in IFS observations from MUSE at the 8.2m VLT of Cerro Paranal.

Two images from the source are identified and, after a careful analysis, confirmed as lensed images coming from the same source.
We modeled the lens with a SIE mass profile and a S\'ersic background source, finding an Einstein radius of $1.45 \pm 0.04$ \arcsec (1.9 kpc), consistent with an isothermal density profile. %This is higher than expected from the velocity dispersion of the lens, indicative of a steep central density profile. 

Independent of our discovery, a recent archival search of public MUSE data by \citet{collier2018} also identified this galaxy as a strong gravitational lens. %Our results are broadly consistent with theirs.
While they predict the dark matter fraction from cosmological hydrodynamical simulations, 
we derive our stellar mass estimate from the SSP synthesis with STARLIGHT to the MUSE data integrated within the Einstein radius.
In both cases, we find a consistent fraction of dark matter within the Einstein radius ($\sim$18 \%).
However, the main difference between both works is the approach used to estimate the average age of the stellar populations in the lens 
%galaxy, which may have dramatical consequences when inferring the initial mass function (IMF) excess parameter ($\alpha$), and led them claim a lightweight IMF rather than a Salpeter IMF for the lens galaxy.}
galaxy. This may have significant consequences when using such parameters to infer the initial mass function (IMF) excess parameter (alpha). Indeed, Collier et al. claim a lightweight IMF rather than a standard Salpeter value for the lens galaxy.

Collier et al. estimate a reference mass-to-light ratio, (M/L)$_{\rm ref}$, for the IMF excess parameter by fitting a single-burst model to a spectrum extracted with an aperture of the Einstein radius.
Following this approach, they are able to recover the dominant population which has a stellar age of 12 Gyr and a metallicity 1.5 Z$_\odot$.
Just considering this single population, they are making the strong/uncertain assumptions that: 
(1) all stars in the galaxy have the same age, metallicity, and were created at the same time; 
(2) the M/L is constant through the extension of the Einstein radius; and 
(3) the IMF is constant through the same extension.
All these assumptions are at odds with current understanding of galaxy formation and evolution from spatially resolved analyses of SPs where each population has its own specific M/L. In addition, these assumptions are clearly not in agreement with the current view of elliptical galaxies (e.g. see  Fig. 5 in \citealt{2018RMxAA..54..217S}, and Fig. 8 in \citealt{2015A&A...581A.103G}).

\begin{figure}
\centering
\includegraphics[trim=0.1cm 0.1cm 0.1cm 0cm,clip=True,width=\columnwidth]{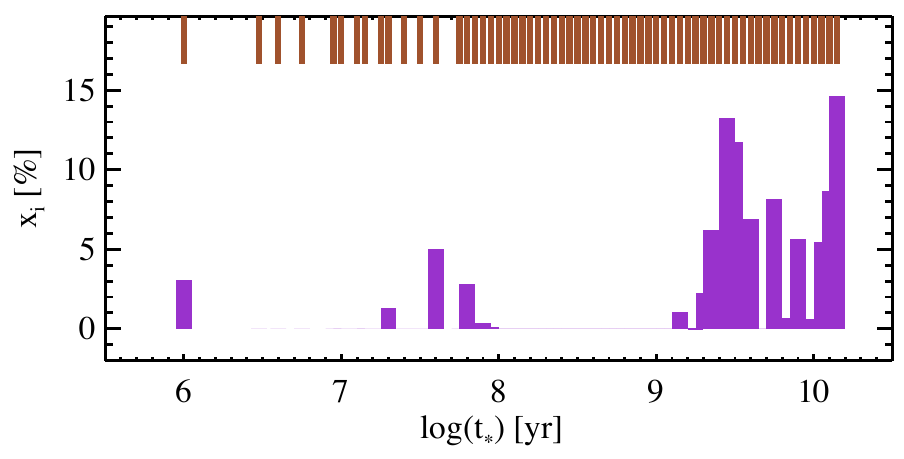}
\caption{Star formation history of the spectrum extracted from the observed MUSE cube with an aperture of the Einstein radius (1.45 arcsec). Vertical brown lines on top correspond to the ages of the base SSPs used in the STARLIGHT fit. The light-weighted average age is 2.6 Gyr. We recover a small fraction of <1 Gyr populations, and two clear peaks around 2.8 and 12.5 Gyr.}
\label{fig:sfh}
\end{figure}

A more detailed approach would consist on performing SSP synthesis to infer the full star formation history. 
The analysis with STARLIGHT on the spectrum of an aperture of the Einstein radius (1.45\arcsec) provides an average age of 2.6 Gyr and metallicity around solar (See Figure \ref{fig:sfh}). We recover SPs of old ages (>10 dex) but a whole range of younger populations, in particular two peaks at 2.8 and 12.5 Gyr (9.4 and 10.1 dex).
As a comparison, Figure 8 in \cite{2005MNRAS.362...41G} shows the distribution of age and metallicity as a function of stellar mass for high signal-to-noise SDSS galaxies. For the stellar mass of the lens galaxy, 10.94 dex, our estimates of age and metallicity are within 1 sigma of the SDSS distribution, while those by Collier et al. are outliers.
In addition, \cite{2004MNRAS.347..691P} in their Figure 2 show that for stellar ages of 12 Gyr the (M/L)$_{\rm ref}$ is around a factor 2 larger than for populations of $\sim$3 Gyr ages. 
Therefore, our extracted values would provide a much lower (M/L)$_{\rm ref}$ and an $\alpha\gtrsim$1.5 totally compatible with a Salpeter IMF.

Finally, we note that to discriminate between different IMFs is unfeasible considering (among other reasons):
(i) the spectral range covered by the current dataset, laking the blue part of the spectrum, in particular de D4000 region;
(ii) our current uncertainties in the SSPs libraries, which make it difficult to measure the M/L at the precision required; 
(iii) not having a good way to measure the [alpha/Fe] gradient which, by itself, could mimic the observed discrepancy; and
(iv) the known uncertainties in the fitting procedures. 
Finally, we stress that putting constraints on the IMF was not the main goal of this work, and we refer to state-of-the-art IMF derivations, such as those from \cite{2015MNRAS.447.1033M} or by the Atlas3D group \citep{2012Natur.484..485C}, to outline the complexity of the matter.

Fifteen other galaxies have been detected in the 1 arcmin$^2$ FoV, in addition to the identification of a possible merger with the lensing galaxy. None of these background sources are strongly lensed.
We provided spectroscopic redshifts and stellar and ionized gas properties for 14 of these background galaxies. 
Object $R$ shows a red continuum, but we were unable to get reliable constrains on its redshift. Deeper observations are needed for such constrains.

In this work we have been able to put constraints on the dark matter content within the Einstein radius of a nearby elliptical galaxy acting as a lens for a background galaxy, and present an analysis in great detail.
Looking to the future, the advent of large surveys such as the Large Synoptic Survey Telescope (LSST) will be able to discover a significant number of new strong-lensed systems, and we demonstrated how wide-field Integral field spectroscopy offers an excellent approach to study them and to precisely model its effects.

%%%%%%%%%%%%%%%%%%%%%%%%%%%%%%%%%%%%%%%%%%%%%%%%%%%%%%%
%%%%%%%%%%%%%%%%%%%%%%%%%%%%%%%%%%%%%%%%%%%%%%%%%%%%%%%
%%%%%%%%%%%%%%%%%%%%%%%%%%%%%%%%%%%%%%%%%%%%%%%%%%%%%%%

\section*{Acknowledgements}

We are greatly thankful to Michael Wood-Vasey, Rachel Mandelbaum, Santiago Gonz\'alez-Gait\'an, Lindsay Oldham, Andrew Zentner, and Jeff Newman, for very valuable input on the early stages of this project. 
L.G. was supported in part by the US National Science Foundation under Grant AST-1311862. 
TEC is funded by a Dennis Sciama Fellowship from the University of Portsmouth.  
Based on observations made with ESO Telescopes at the La Silla Paranal Observatory under program 98.D-0115(A).
Llenya d'alzina, vi de sarment, oli d'oliva, pa de forment.

%%%%%%%%%%%%%%%%%%%%%%%%%%%%%%%%%%%%%%%%%%%%%%%%%%

\bibliographystyle{mnras}
\bibliography{bib}

\begin{thebibliography}{}
\makeatletter
\relax
\def\mn@urlcharsother{\let\do\@makeother \do\$\do\&\do\#\do\^\do\_\do\%\do\~}
\def\mn@doi{\begingroup\mn@urlcharsother \@ifnextchar [ {\mn@doi@}
  {\mn@doi@[]}}
\def\mn@doi@[#1]#2{\def\@tempa{#1}\ifx\@tempa\@empty \href
  {http://dx.doi.org/#2} {doi:#2}\else \href {http://dx.doi.org/#2} {#1}\fi
  \endgroup}
\def\mn@eprint#1#2{\mn@eprint@#1:#2::\@nil}
\def\mn@eprint@arXiv#1{\href {http://arxiv.org/abs/#1} {{\tt arXiv:#1}}}
\def\mn@eprint@dblp#1{\href {http://dblp.uni-trier.de/rec/bibtex/#1.xml}
  {dblp:#1}}
\def\mn@eprint@#1:#2:#3:#4\@nil{\def\@tempa {#1}\def\@tempb {#2}\def\@tempc
  {#3}\ifx \@tempc \@empty \let \@tempc \@tempb \let \@tempb \@tempa \fi \ifx
  \@tempb \@empty \def\@tempb {arXiv}\fi \@ifundefined
  {mn@eprint@\@tempb}{\@tempb:\@tempc}{\expandafter \expandafter \csname
  mn@eprint@\@tempb\endcsname \expandafter{\@tempc}}}

\bibitem[\protect\citeauthoryear{{Auger}, {Treu}, {Bolton}, {Gavazzi},
  {Koopmans}, {Marshall}, {Moustakas}  \& {Burles}}{{Auger}
  et~al.}{2010}]{2010ApJ...724..511A}
{Auger} M.~W.,  {Treu} T.,  {Bolton} A.~S.,  {Gavazzi} R.,  {Koopmans}
  L.~V.~E.,  {Marshall} P.~J.,  {Moustakas} L.~A.,   {Burles} S.,  2010,
  \mn@doi [\apj] {10.1088/0004-637X/724/1/511}, \href
  {http://adsabs.harvard.edu/abs/2010ApJ...724..511A} {724, 511}

\bibitem[\protect\citeauthoryear{{Bacon} et~al.,}{{Bacon}
  et~al.}{2010}]{2010SPIE.7735E..08B}
{Bacon} R.,  et~al., 2010, in Ground-based and Airborne Instrumentation for
  Astronomy III. p. 773508, \mn@doi{10.1117/12.856027}

\bibitem[\protect\citeauthoryear{{Bekki}, {Couch}  \& {Shioya}}{{Bekki}
  et~al.}{2002}]{bekki02}
{Bekki} K.,  {Couch} W.~J.,   {Shioya} Y.,  2002, \mn@doi [\apj]
  {10.1086/342221}, \href {http://adsabs.harvard.edu/abs/2002ApJ...577..651B}
  {577, 651}

\bibitem[\protect\citeauthoryear{{Belfiore}, {Maiolino}  \&
  {Bothwell}}{{Belfiore} et~al.}{2016}]{belf16a}
{Belfiore} F.,  {Maiolino} R.,   {Bothwell} M.,  2016, \mn@doi [\mnras]
  {10.1093/mnras/stv2332}, \href
  {http://adsabs.harvard.edu/abs/2016MNRAS.455.1218B} {455, 1218}

\bibitem[\protect\citeauthoryear{{Bertin} et~al.,}{{Bertin}
  et~al.}{1994}]{1994A&A...292..381B}
{Bertin} G.,  et~al., 1994, \aap, \href
  {http://adsabs.harvard.edu/abs/1994A%26A...292..381B} {292, 381}

\bibitem[\protect\citeauthoryear{{Binette}, {Magris}, {Stasi{\'n}ska}  \&
  {Bruzual}}{{Binette} et~al.}{1994}]{binn94}
{Binette} L.,  {Magris} C.~G.,  {Stasi{\'n}ska} G.,   {Bruzual} A.~G.,  1994,
  \aap, \href {http://adsabs.harvard.edu/abs/1994A%26A...292...13B} {292, 13}

\bibitem[\protect\citeauthoryear{{Blakeslee} et~al.,}{{Blakeslee}
  et~al.}{2004}]{2004ApJ...602L...9B}
{Blakeslee} J.~P.,  et~al., 2004, \mn@doi [\apjl] {10.1086/382505}, \href
  {http://adsabs.harvard.edu/abs/2004ApJ...602L...9B} {602, L9}

\bibitem[\protect\citeauthoryear{{Bolton}, {Burles}, {Koopmans}, {Treu}  \&
  {Moustakas}}{{Bolton} et~al.}{2006}]{2006ApJ...638..703B}
{Bolton} A.~S.,  {Burles} S.,  {Koopmans} L.~V.~E.,  {Treu} T.,   {Moustakas}
  L.~A.,  2006, \mn@doi [\apj] {10.1086/498884}, \href
  {http://adsabs.harvard.edu/abs/2006ApJ...638..703B} {638, 703}

\bibitem[\protect\citeauthoryear{{Bonvin} et~al.,}{{Bonvin}
  et~al.}{2017}]{bonvin2017}
{Bonvin} V.,  et~al., 2017, \mn@doi [\mnras] {10.1093/mnras/stw3006}, \href
  {http://adsabs.harvard.edu/abs/2017MNRAS.465.4914B} {465, 4914}

\bibitem[\protect\citeauthoryear{{Bundy} et~al.,}{{Bundy}
  et~al.}{2015}]{Bundy15}
{Bundy} K.,  et~al., 2015, \mn@doi [\apj] {10.1088/0004-637X/798/1/7}, \href
  {http://adsabs.harvard.edu/abs/2015ApJ...798....7B} {798, 7}

\bibitem[\protect\citeauthoryear{{Cappellari} et~al.,}{{Cappellari}
  et~al.}{2012}]{2012Natur.484..485C}
{Cappellari} M.,  et~al., 2012, \mn@doi [\nat] {10.1038/nature10972}, \href
  {http://adsabs.harvard.edu/abs/2012Natur.484..485C} {484, 485}

\bibitem[\protect\citeauthoryear{{Cappellari} et~al.,}{{Cappellari}
  et~al.}{2013}]{2013MNRAS.432.1709C}
{Cappellari} M.,  et~al., 2013, \mn@doi [\mnras] {10.1093/mnras/stt562}, \href
  {http://adsabs.harvard.edu/abs/2013MNRAS.432.1709C} {432, 1709}

\bibitem[\protect\citeauthoryear{{Cappellari} et~al.,}{{Cappellari}
  et~al.}{2015}]{2015ApJ...804L..21C}
{Cappellari} M.,  et~al., 2015, \mn@doi [\apjl] {10.1088/2041-8205/804/1/L21},
  \href {http://adsabs.harvard.edu/abs/2015ApJ...804L..21C} {804, L21}

\bibitem[\protect\citeauthoryear{{Cardelli}, {Clayton}  \& {Mathis}}{{Cardelli}
  et~al.}{1989}]{1989ApJ...345..245C}
{Cardelli} J.~A.,  {Clayton} G.~C.,   {Mathis} J.~S.,  1989, \mn@doi [\apj]
  {10.1086/167900}, \href {http://adsabs.harvard.edu/abs/1989ApJ...345..245C}
  {345, 245}

\bibitem[\protect\citeauthoryear{{Charbonnel}, {Meynet}, {Maeder}, {Schaller}
  \& {Schaerer}}{{Charbonnel} et~al.}{1993}]{1993A&AS..101..415C}
{Charbonnel} C.,  {Meynet} G.,  {Maeder} A.,  {Schaller} G.,   {Schaerer} D.,
  1993, \aaps, \href {http://adsabs.harvard.edu/abs/1993A%26AS..101..415C}
  {101, 415}

\bibitem[\protect\citeauthoryear{{Cheung} et~al.,}{{Cheung}
  et~al.}{2016}]{Cheung16}
{Cheung} E.,  et~al., 2016, \mn@doi [\nat] {10.1038/nature18006}, \href
  {http://adsabs.harvard.edu/abs/2016Natur.533..504C} {533, 504}

\bibitem[\protect\citeauthoryear{{Cid Fernandes}, {Mateus}, {Sodr{\'e}},
  {Stasi{\'n}ska}  \& {Gomes}}{{Cid Fernandes}
  et~al.}{2005}]{2005MNRAS.358..363C}
{Cid Fernandes} R.,  {Mateus} A.,  {Sodr{\'e}} L.,  {Stasi{\'n}ska} G.,
  {Gomes} J.~M.,  2005, \mn@doi [\mnras] {10.1111/j.1365-2966.2005.08752.x},
  \href {http://adsabs.harvard.edu/abs/2005MNRAS.358..363C} {358, 363}

\bibitem[\protect\citeauthoryear{{Collett} \& {Auger}}{{Collett} \&
  {Auger}}{2014}]{2014MNRAS.443..969C}
{Collett} T.~E.,  {Auger} M.~W.,  2014, \mn@doi [\mnras]
  {10.1093/mnras/stu1190}, \href
  {http://adsabs.harvard.edu/abs/2014MNRAS.443..969C} {443, 969}

\bibitem[\protect\citeauthoryear{{Collier}, {Smith}  \& {Lucey}}{{Collier}
  et~al.}{2018}]{collier2018}
{Collier} W.~P.,  {Smith} R.~J.,   {Lucey} J.~R.,  2018, preprint, \href
  {http://adsabs.harvard.edu/abs/2018arXiv180307082C} {} (\mn@eprint {arXiv}
  {1803.07082})

\bibitem[\protect\citeauthoryear{{Einstein}}{{Einstein}}{1915}]{1915SPAW.......844E}
{Einstein} A.,  1915, Sitzungsberichte der K{\"o}niglich Preu{\ss}ischen
  Akademie der Wissenschaften (Berlin), Seite 844-847., \href
  {http://adsabs.harvard.edu/abs/1915SPAW.......844E} {}

\bibitem[\protect\citeauthoryear{{Falc{\'o}n-Barroso},
  {S{\'a}nchez-Bl{\'a}zquez}, {Vazdekis}, {Ricciardelli}, {Cardiel}, {Cenarro},
  {Gorgas}  \& {Peletier}}{{Falc{\'o}n-Barroso}
  et~al.}{2011}]{2011A&A...532A..95F}
{Falc{\'o}n-Barroso} J.,  {S{\'a}nchez-Bl{\'a}zquez} P.,  {Vazdekis} A.,
  {Ricciardelli} E.,  {Cardiel} N.,  {Cenarro} A.~J.,  {Gorgas} J.,
  {Peletier} R.~F.,  2011, \mn@doi [\aap] {10.1051/0004-6361/201116842}, \href
  {http://adsabs.harvard.edu/abs/2011A%26A...532A..95F} {532, A95}

\bibitem[\protect\citeauthoryear{{Galbany} et~al.,}{{Galbany}
  et~al.}{2014}]{2014A&A...572A..38G}
{Galbany} L.,  et~al., 2014, \mn@doi [\aap] {10.1051/0004-6361/201424717},
  \href {http://adsabs.harvard.edu/abs/2014A%26A...572A..38G} {572, A38}

\bibitem[\protect\citeauthoryear{{Galbany} et~al.,}{{Galbany}
  et~al.}{2016a}]{2016MNRAS.455.4087G}
{Galbany} L.,  et~al., 2016a, \mn@doi [\mnras] {10.1093/mnras/stv2620}, \href
  {http://adsabs.harvard.edu/abs/2016MNRAS.455.4087G} {455, 4087}

\bibitem[\protect\citeauthoryear{{Galbany} et~al.,}{{Galbany}
  et~al.}{2016b}]{2016A&A...591A..48G}
{Galbany} L.,  et~al., 2016b, \mn@doi [\aap] {10.1051/0004-6361/201528045},
  \href {http://adsabs.harvard.edu/abs/2016A%26A...591A..48G} {591, A48}

\bibitem[\protect\citeauthoryear{{Galbany} et~al.,}{{Galbany}
  et~al.}{2018}]{2018ApJ...855..107G}
{Galbany} L.,  et~al., 2018, \mn@doi [\apj] {10.3847/1538-4357/aaaf20}, \href
  {http://adsabs.harvard.edu/abs/2018ApJ...855..107G} {855, 107}

\bibitem[\protect\citeauthoryear{{Gallazzi}, {Charlot}, {Brinchmann}, {White}
  \& {Tremonti}}{{Gallazzi} et~al.}{2005}]{2005MNRAS.362...41G}
{Gallazzi} A.,  {Charlot} S.,  {Brinchmann} J.,  {White} S.~D.~M.,   {Tremonti}
  C.~A.,  2005, \mn@doi [\mnras] {10.1111/j.1365-2966.2005.09321.x}, \href
  {http://adsabs.harvard.edu/abs/2005MNRAS.362...41G} {362, 41}

\bibitem[\protect\citeauthoryear{{Garc{\'{\i}}a-Lorenzo}, {S{\'a}nchez},
  {Mediavilla}, {Gonz{\'a}lez-Serrano}  \&
  {Christensen}}{{Garc{\'{\i}}a-Lorenzo} et~al.}{2005}]{bego05}
{Garc{\'{\i}}a-Lorenzo} B.,  {S{\'a}nchez} S.~F.,  {Mediavilla} E.,
  {Gonz{\'a}lez-Serrano} J.~I.,   {Christensen} L.,  2005, \mn@doi [\apj]
  {10.1086/427429}, \href {http://adsabs.harvard.edu/abs/2005ApJ...621..146G}
  {621, 146}

\bibitem[\protect\citeauthoryear{{Gavazzi}, {Marshall}, {Treu}  \&
  {Sonnenfeld}}{{Gavazzi} et~al.}{2014}]{2014ApJ...785..144G}
{Gavazzi} R.,  {Marshall} P.~J.,  {Treu} T.,   {Sonnenfeld} A.,  2014, \mn@doi
  [\apj] {10.1088/0004-637X/785/2/144}, \href
  {http://adsabs.harvard.edu/abs/2014ApJ...785..144G} {785, 144}

\bibitem[\protect\citeauthoryear{{Girardi}, {Bressan}, {Bertelli}  \&
  {Chiosi}}{{Girardi} et~al.}{2000}]{2000A&AS..141..371G}
{Girardi} L.,  {Bressan} A.,  {Bertelli} G.,   {Chiosi} C.,  2000, \mn@doi
  [\aaps] {10.1051/aas:2000126}, \href
  {http://adsabs.harvard.edu/abs/2000A%26AS..141..371G} {141, 371}

\bibitem[\protect\citeauthoryear{{Gomes} et~al.,}{{Gomes}
  et~al.}{2016a}]{Gomes16b}
{Gomes} J.~M.,  et~al., 2016a, \mn@doi [\aap] {10.1051/0004-6361/201525974},
  \href {http://adsabs.harvard.edu/abs/2016A%26A...585A..92G} {585, A92}

\bibitem[\protect\citeauthoryear{{Gomes} et~al.,}{{Gomes}
  et~al.}{2016b}]{Gomes16a}
{Gomes} J.~M.,  et~al., 2016b, \mn@doi [\aap] {10.1051/0004-6361/201525976},
  \href {http://adsabs.harvard.edu/abs/2016A%26A...588A..68G} {588, A68}

\bibitem[\protect\citeauthoryear{{Gonz{\'a}lez Delgado}, {Cervi{\~n}o},
  {Martins}, {Leitherer}  \& {Hauschildt}}{{Gonz{\'a}lez Delgado}
  et~al.}{2005}]{2005MNRAS.357..945G}
{Gonz{\'a}lez Delgado} R.~M.,  {Cervi{\~n}o} M.,  {Martins} L.~P.,  {Leitherer}
  C.,   {Hauschildt} P.~H.,  2005, \mn@doi [\mnras]
  {10.1111/j.1365-2966.2005.08692.x}, \href
  {http://adsabs.harvard.edu/abs/2005MNRAS.357..945G} {357, 945}

\bibitem[\protect\citeauthoryear{{Gonz{\'a}lez Delgado} et~al.,}{{Gonz{\'a}lez
  Delgado} et~al.}{2015}]{2015A&A...581A.103G}
{Gonz{\'a}lez Delgado} R.~M.,  et~al., 2015, \mn@doi [\aap]
  {10.1051/0004-6361/201525938}, \href
  {http://adsabs.harvard.edu/abs/2015A%26A...581A.103G} {581, A103}

\bibitem[\protect\citeauthoryear{{Jones} et~al.,}{{Jones}
  et~al.}{2009}]{2009MNRAS.399..683J}
{Jones} D.~H.,  et~al., 2009, \mn@doi [\mnras]
  {10.1111/j.1365-2966.2009.15338.x}, \href
  {http://adsabs.harvard.edu/abs/2009MNRAS.399..683J} {399, 683}

\bibitem[\protect\citeauthoryear{{Kauffmann} et~al.,}{{Kauffmann}
  et~al.}{2003}]{2003MNRAS.346.1055K}
{Kauffmann} G.,  et~al., 2003, \mn@doi [\mnras]
  {10.1111/j.1365-2966.2003.07154.x}, \href
  {http://adsabs.harvard.edu/abs/2003MNRAS.346.1055K} {346, 1055}

\bibitem[\protect\citeauthoryear{{Kehrig} et~al.,}{{Kehrig}
  et~al.}{2012}]{kehrig12}
{Kehrig} C.,  et~al., 2012, \mn@doi [\aap] {10.1051/0004-6361/201118357}, \href
  {http://adsabs.harvard.edu/abs/2012A%26A...540A..11K} {540, A11}

\bibitem[\protect\citeauthoryear{{Kennicutt}}{{Kennicutt}}{1998}]{1998ARA&A..36..189K}
{Kennicutt} Jr. R.~C.,  1998, \mn@doi [\araa] {10.1146/annurev.astro.36.1.189},
  \href {http://adsabs.harvard.edu/abs/1998ARA%26A..36..189K} {36, 189}

\bibitem[\protect\citeauthoryear{{Kewley}, {Dopita}, {Sutherland}, {Heisler}
  \& {Trevena}}{{Kewley} et~al.}{2001}]{2001ApJ...556..121K}
{Kewley} L.~J.,  {Dopita} M.~A.,  {Sutherland} R.~S.,  {Heisler} C.~A.,
  {Trevena} J.,  2001, \mn@doi [\apj] {10.1086/321545}, \href
  {http://adsabs.harvard.edu/abs/2001ApJ...556..121K} {556, 121}

\bibitem[\protect\citeauthoryear{{Kr{\"u}hler}, {Kuncarayakti}, {Schady},
  {Anderson}, {Galbany}  \& {Gensior}}{{Kr{\"u}hler}
  et~al.}{2017}]{2017A&A...602A..85K}
{Kr{\"u}hler} T.,  {Kuncarayakti} H.,  {Schady} P.,  {Anderson} J.~P.,
  {Galbany} L.,   {Gensior} J.,  2017, \mn@doi [\aap]
  {10.1051/0004-6361/201630268}, \href
  {http://adsabs.harvard.edu/abs/2017A%26A...602A..85K} {602, A85}

\bibitem[\protect\citeauthoryear{{Lanusse}, {Ma}, {Li}, {Collett}, {Li},
  {Ravanbakhsh}, {Mandelbaum}  \& {P{\'o}czos}}{{Lanusse}
  et~al.}{2018}]{2018MNRAS.473.3895L}
{Lanusse} F.,  {Ma} Q.,  {Li} N.,  {Collett} T.~E.,  {Li} C.-L.,  {Ravanbakhsh}
  S.,  {Mandelbaum} R.,   {P{\'o}czos} B.,  2018, \mn@doi [\mnras]
  {10.1093/mnras/stx1665}, \href
  {http://adsabs.harvard.edu/abs/2018MNRAS.473.3895L} {473, 3895}

\bibitem[\protect\citeauthoryear{{Marino} et~al.,}{{Marino}
  et~al.}{2013}]{2013A&A...559A.114M}
{Marino} R.~A.,  et~al., 2013, \mn@doi [\aap] {10.1051/0004-6361/201321956},
  \href {http://adsabs.harvard.edu/abs/2013A%26A...559A.114M} {559, A114}

\bibitem[\protect\citeauthoryear{{Mart{\'{\i}}n-Navarro}, {La Barbera},
  {Vazdekis}, {Falc{\'o}n-Barroso}  \& {Ferreras}}{{Mart{\'{\i}}n-Navarro}
  et~al.}{2015}]{2015MNRAS.447.1033M}
{Mart{\'{\i}}n-Navarro} I.,  {La Barbera} F.,  {Vazdekis} A.,
  {Falc{\'o}n-Barroso} J.,   {Ferreras} I.,  2015, \mn@doi [\mnras]
  {10.1093/mnras/stu2480}, \href
  {http://adsabs.harvard.edu/abs/2015MNRAS.447.1033M} {447, 1033}

\bibitem[\protect\citeauthoryear{{M{\'e}ndez-Abreu}, {Aguerri}, {Corsini}  \&
  {Simonneau}}{{M{\'e}ndez-Abreu} et~al.}{2008}]{mendezabreu08}
{M{\'e}ndez-Abreu} J.,  {Aguerri} J.~A.~L.,  {Corsini} E.~M.,   {Simonneau} E.,
   2008, \mn@doi [\aap] {10.1051/0004-6361:20078089}, \href
  {http://adsabs.harvard.edu/abs/2008A%26A...478..353M} {478, 353}

\bibitem[\protect\citeauthoryear{{M{\'e}ndez-Abreu}, {Debattista}, {Corsini}
  \& {Aguerri}}{{M{\'e}ndez-Abreu} et~al.}{2014}]{mendezabreu14}
{M{\'e}ndez-Abreu} J.,  {Debattista} V.~P.,  {Corsini} E.~M.,   {Aguerri}
  J.~A.~L.,  2014, \mn@doi [\aap] {10.1051/0004-6361/201423955}, \href
  {http://adsabs.harvard.edu/abs/2014A%26A...572A..25M} {572, A25}

\bibitem[\protect\citeauthoryear{{Narayan} \& {Bartelmann}}{{Narayan} \&
  {Bartelmann}}{1996}]{1996astro.ph..6001N}
{Narayan} R.,  {Bartelmann} M.,  1996, ArXiv Astrophysics e-prints, \href
  {http://adsabs.harvard.edu/abs/1996astro.ph..6001N} {}

\bibitem[\protect\citeauthoryear{{Newman}, {Smith}, {Conroy}, {Villaume}  \&
  {van Dokkum}}{{Newman} et~al.}{2017}]{2017ApJ...845..157N}
{Newman} A.~B.,  {Smith} R.~J.,  {Conroy} C.,  {Villaume} A.,   {van Dokkum}
  P.,  2017, \mn@doi [\apj] {10.3847/1538-4357/aa816d}, \href
  {http://adsabs.harvard.edu/abs/2017ApJ...845..157N} {845, 157}

\bibitem[\protect\citeauthoryear{{Papaderos} et~al.,}{{Papaderos}
  et~al.}{2013}]{papa13}
{Papaderos} P.,  et~al., 2013, preprint, \href
  {http://adsabs.harvard.edu/abs/2013arXiv1306.2338P} {} (\mn@eprint {arXiv}
  {1306.2338})

\bibitem[\protect\citeauthoryear{{Planck Collaboration} et~al.,}{{Planck
  Collaboration} et~al.}{2016}]{2016A&A...594A..13P}
{Planck Collaboration} et~al., 2016, \mn@doi [\aap]
  {10.1051/0004-6361/201525830}, \href
  {http://adsabs.harvard.edu/abs/2016A%26A...594A..13P} {594, A13}

\bibitem[\protect\citeauthoryear{{Portinari}, {Sommer-Larsen}  \&
  {Tantalo}}{{Portinari} et~al.}{2004}]{2004MNRAS.347..691P}
{Portinari} L.,  {Sommer-Larsen} J.,   {Tantalo} R.,  2004, \mn@doi [\mnras]
  {10.1111/j.1365-2966.2004.07207.x}, \href
  {http://adsabs.harvard.edu/abs/2004MNRAS.347..691P} {347, 691}

\bibitem[\protect\citeauthoryear{{Salpeter}}{{Salpeter}}{1955}]{1955ApJ...121..161S}
{Salpeter} E.~E.,  1955, \mn@doi [\apj] {10.1086/145971}, \href
  {http://adsabs.harvard.edu/abs/1955ApJ...121..161S} {121, 161}

\bibitem[\protect\citeauthoryear{{S{\'a}nchez} et~al.,}{{S{\'a}nchez}
  et~al.}{2016}]{2016RMxAA..52...21S}
{S{\'a}nchez} S.~F.,  et~al., 2016, \rmxaa, \href
  {http://adsabs.harvard.edu/abs/2016RMxAA..52...21S} {52, 21}

\bibitem[\protect\citeauthoryear{{Sanchez} et~al.,}{{Sanchez}
  et~al.}{2017}]{sanchez17b}
{Sanchez} S.~F.,  et~al., 2017, preprint, \href
  {http://adsabs.harvard.edu/abs/2017arXiv170905438S} {} (\mn@eprint {arXiv}
  {1709.05438})

\bibitem[\protect\citeauthoryear{{S{\'a}nchez} et~al.,}{{S{\'a}nchez}
  et~al.}{2018}]{2018RMxAA..54..217S}
{S{\'a}nchez} S.~F.,  et~al., 2018, \rmxaa, \href
  {http://adsabs.harvard.edu/abs/2018RMxAA..54..217S} {54, 217}

\bibitem[\protect\citeauthoryear{{Sarzi} et~al.,}{{Sarzi}
  et~al.}{2010}]{sarzi10}
{Sarzi} M.,  et~al., 2010, \mn@doi [\mnras] {10.1111/j.1365-2966.2009.16039.x},
  \href {http://adsabs.harvard.edu/abs/2010MNRAS.402.2187S} {402, 2187}

\bibitem[\protect\citeauthoryear{{Schaerer}, {Meynet}, {Maeder}  \&
  {Schaller}}{{Schaerer} et~al.}{1993a}]{1993A&AS...98..523S}
{Schaerer} D.,  {Meynet} G.,  {Maeder} A.,   {Schaller} G.,  1993a, \aaps,
  \href {http://adsabs.harvard.edu/abs/1993A%26AS...98..523S} {98, 523}

\bibitem[\protect\citeauthoryear{{Schaerer}, {Charbonnel}, {Meynet}, {Maeder}
  \& {Schaller}}{{Schaerer} et~al.}{1993b}]{1993A&AS..102..339S}
{Schaerer} D.,  {Charbonnel} C.,  {Meynet} G.,  {Maeder} A.,   {Schaller} G.,
  1993b, \aaps, \href {http://adsabs.harvard.edu/abs/1993A%26AS..102..339S}
  {102, 339}

\bibitem[\protect\citeauthoryear{{Schaller}, {Schaerer}, {Meynet}  \&
  {Maeder}}{{Schaller} et~al.}{1992}]{1992A&AS...96..269S}
{Schaller} G.,  {Schaerer} D.,  {Meynet} G.,   {Maeder} A.,  1992, \aaps, \href
  {http://adsabs.harvard.edu/abs/1992A%26AS...96..269S} {96, 269}

\bibitem[\protect\citeauthoryear{{Schlafly} \& {Finkbeiner}}{{Schlafly} \&
  {Finkbeiner}}{2011}]{2011ApJ...737..103S}
{Schlafly} E.~F.,  {Finkbeiner} D.~P.,  2011, \mn@doi [\apj]
  {10.1088/0004-637X/737/2/103}, \href
  {http://adsabs.harvard.edu/abs/2011ApJ...737..103S} {737, 103}

\bibitem[\protect\citeauthoryear{{Singh} et~al.,}{{Singh}
  et~al.}{2013}]{sign13}
{Singh} R.,  et~al., 2013, \mn@doi [\aap] {10.1051/0004-6361/201322062}, \href
  {http://adsabs.harvard.edu/abs/2013A%26A...558A..43S} {558, A43}

\bibitem[\protect\citeauthoryear{{Smith}}{{Smith}}{2017}]{2017MNRAS.464L..46S}
{Smith} R.~J.,  2017, \mn@doi [\mnras] {10.1093/mnrasl/slw174}, \href
  {http://adsabs.harvard.edu/abs/2017MNRAS.464L..46S} {464, L46}

\bibitem[\protect\citeauthoryear{{Smith}, {Blakeslee}, {Lucey}  \&
  {Tonry}}{{Smith} et~al.}{2005}]{2005ApJ...625L.103S}
{Smith} R.~J.,  {Blakeslee} J.~P.,  {Lucey} J.~R.,   {Tonry} J.,  2005, \mn@doi
  [\apjl] {10.1086/431240}, \href
  {http://adsabs.harvard.edu/abs/2005ApJ...625L.103S} {625, L103}

\bibitem[\protect\citeauthoryear{{Smith}, {Lucey}  \& {Conroy}}{{Smith}
  et~al.}{2015}]{2015MNRAS.449.3441S}
{Smith} R.~J.,  {Lucey} J.~R.,   {Conroy} C.,  2015, \mn@doi [\mnras]
  {10.1093/mnras/stv518}, \href
  {http://adsabs.harvard.edu/abs/2015MNRAS.449.3441S} {449, 3441}

\bibitem[\protect\citeauthoryear{{Stasi{\'n}ska} et~al.,}{{Stasi{\'n}ska}
  et~al.}{2008}]{sta08}
{Stasi{\'n}ska} G.,  et~al., 2008, \mn@doi [\mnras]
  {10.1111/j.1745-3933.2008.00550.x}, \href
  {http://adsabs.harvard.edu/abs/2008MNRAS.391L..29S} {391, L29}

\bibitem[\protect\citeauthoryear{{Talbot} et~al.,}{{Talbot}
  et~al.}{2018}]{talbot2018}
{Talbot} M.~S.,  et~al., 2018, \mn@doi [\mnras] {10.1093/mnras/sty653}, \href
  {http://adsabs.harvard.edu/abs/2018MNRAS.tmp..630T} {}

\bibitem[\protect\citeauthoryear{{Treu}}{{Treu}}{2010}]{2010ARA&A..48...87T}
{Treu} T.,  2010, \mn@doi [\araa] {10.1146/annurev-astro-081309-130924}, \href
  {http://adsabs.harvard.edu/abs/2010ARA%26A..48...87T} {48, 87}

\bibitem[\protect\citeauthoryear{{Vazdekis}, {S{\'a}nchez-Bl{\'a}zquez},
  {Falc{\'o}n-Barroso}, {Cenarro}, {Beasley}, {Cardiel}, {Gorgas}  \&
  {Peletier}}{{Vazdekis} et~al.}{2010}]{2010MNRAS.404.1639V}
{Vazdekis} A.,  {S{\'a}nchez-Bl{\'a}zquez} P.,  {Falc{\'o}n-Barroso} J.,
  {Cenarro} A.~J.,  {Beasley} M.~A.,  {Cardiel} N.,  {Gorgas} J.,   {Peletier}
  R.~F.,  2010, \mn@doi [\mnras] {10.1111/j.1365-2966.2010.16407.x}, \href
  {http://adsabs.harvard.edu/abs/2010MNRAS.404.1639V} {404, 1639}

\bibitem[\protect\citeauthoryear{{Walker} et~al.,}{{Walker}
  et~al.}{2013}]{2013ATel.5567....1W}
{Walker} E.~S.,  et~al., 2013, The Astronomer's Telegram, \href
  {http://adsabs.harvard.edu/abs/2013ATel.5567....1W} {5567}

\bibitem[\protect\citeauthoryear{{Zwicky}}{{Zwicky}}{1937}]{1937PhRv...51..290Z}
{Zwicky} F.,  1937, \mn@doi [Physical Review] {10.1103/PhysRev.51.290}, \href
  {http://adsabs.harvard.edu/abs/1937PhRv...51..290Z} {51, 290}

\makeatother
\end{thebibliography}

\begin{table*}\scriptsize
\caption{Extinction corrected fluxes of the most prominent gas phase emission lines. All fluxes in units of 10$^{-18}$ erg s$^{-1}$ cm$^{-2}$ $\AA^{-1}$. }
\label{tab:otherelines}
\begin{center}
\begin{tabular}{cccccccccccc}
\hline\hline
Galaxy &
[O\,II]\,$\lambda\lambda$3727,29 & 
H\,$\delta$ & 
H\,$\gamma$ & 
H\,$\beta$ & 
[O\,III]\,$\lambda$4959 & 
[O\,III]\,$\lambda$5007 & 
[N\,II]\,$\lambda$6548 & 
H\,$\alpha$ & 
[N\,II] \,$\lambda$6583 &
[S\,II]\,$\lambda$6717 & 
[S\,II]\,$\lambda$6731 \\
\hline
D & $-$ & $-$ & $-$ & $-$ & $-$ & $-$ & $-$ & $-$ & $-$ & $-$ & $-$ \\
E & $-$ & $-$ & $-$ & $-$ & $-$ & $-$ & $-$ & $-$ & $-$ & $-$ & $-$ \\
F & 26.89 (7.67) & $-$ & $-$ & 12.19 (5.94) & 22.04 (9.27) & 76.08 (27.40) & $-$ & $-$ & $-$ & $-$ & $-$ \\
G & $-$ & $-$ & $-$ & $-$ & $-$ & $-$ & $-$ & $-$ & $-$ & $-$ & $-$ \\
H & 96.54 (21.82) & $-$ & $-$ & 28.70 (11.09) & 25.91 (9.88) & 71.83 (24.28) & $-$ & $-$ & $-$ & $-$ & $-$ \\
I & $-$ & $-$ & $-$ & $-$ & $-$ & $-$ & $-$ & $-$ & $-$ & $-$ & $-$ \\
J & $-$ & $-$ & $-$ & $-$ & $-$ & $-$ & $-$ & $-$ & $-$ & $-$ & $-$ \\
K & $-$ & $-$ & $-$ & 57.56 (20.42) & 29.46 (11.50) & 84.40 (31.79) & $-$ & 164.62 (61.47) & $-$ & 164.62 (61.47) & $-$ \\
L & $-$ & $-$ & $-$ & 236.77 (86.09) & 416.37 (154.50) & 1256.91 (472.34) & $-$ & 677.17 (251.72) & $-$ & 677.17 (251.72) & $-$ \\
M & $-$ & $-$ & $-$ & $-$ & $-$ & $-$ & $-$ & $-$ & $-$ & $-$ & $-$ \\
N & 26.39 (3.69) & $-$ & $-$ & $-$ & $-$ & $-$ & $-$ & $-$ & $-$ & $-$ & $-$ \\
O & 24.03 (4.67) & $-$ & $-$ & $-$ & $-$ & $-$ & $-$ & $-$ & $-$ & $-$ & $-$ \\
P & 47.30 (9.43) & $-$ & 9.79 (2.38) & $-$ & $-$ & 16.58 (3.66) & $-$ & $-$ & $-$ & $-$ & $-$ \\
Q & 59.05 (12.89) & $-$ & $-$ & 20.78 (4.70) & $-$ & $-$ & $-$ & $-$ & $-$ & $-$ & $-$ \\
\hline
\end{tabular}
\end{center}
\end{table*}

\begin{table*}\footnotesize
\caption{Properties extracted from aperture spectra of all sources in the residual cube after subtracting the lens model, except for the lens which was extracted in the observed cube.}
\label{tab:otherprop}
\begin{center}
\begin{tabular}{cccccccc}
\hline\hline
Galaxy &
A$_V^{\rm gas}$ &
12+log$_{10}$ (O/H) &
log$_{10}$ (SFR [M$_\odot$ yr$^{-1}$]) &
log$_{10}$ (M [M$_\odot$]) &
A$_V^*$ &
<log$_{10}$ (t$_*$ [yr])> &
<log$_{10}$ Z$_*$> \\
 & [mag] & [dex] & [dex] & [dex] & [mag] & [dex] & [dex] \\
\hline
D &       $-$   &    $-$      &   $-$    & 10.63 & 0.00 & 9.73 (0.35) & 0.027 (0.007) \\
E &       $-$   &    $-$      &   $-$    & 11.11 & 0.17 & 9.84 (0.33) & 0.023 (0.014) \\
F &       $-$   &    $-$      &   $-$    &   $-$ &  $-$ &      $-$    &       $-$     \\
G &       $-$   &    $-$      &   $-$    & 10.27 & 0.10 & 9.53 (0.52) & 0.024 (0.014) \\
H &       $-$   &    $-$      &   $-$    &   $-$ &  $-$ &      $-$    &       $-$     \\
I &       $-$   &    $-$      &   $-$    & 11.18 & 0.00 & 9.64 (0.33) & 0.032 (0.000) \\
J &       $-$   &    $-$      &   $-$    & 10.75 & 0.00 & 9.64 (0.38) & 0.031 (0.003) \\
K & 0.31 (0.03) &    $-$      & 0.031526 &   $-$ &  $-$ &      $-$    &       $-$     \\
L & 0.05 (0.02) &    $-$      & 0.129686 &   $-$ &  $-$ &      $-$    &       $-$     \\
M &       $-$   &    $-$      &   $-$    & 10.89 & 0.00 & 9.85 (0.35) & 0.032 (0.000) \\
N &       $-$   &    $-$      &   $-$    & 11.06 & 0.09 &10.12 (0.08) & 0.019 (0.000) \\
O &       $-$   &    $-$      &   $-$    &  9.91 & 0.05 & 8.98 (0.33) & 0.023 (0.008) \\
P &       $-$   &    $-$      &   $-$    &   $-$ &  $-$ &      $-$    &       $-$     \\
Q &       $-$   &    $-$      &   $-$    &   $-$ &  $-$ &      $-$    &       $-$     \\
R &       $-$   &    $-$      &   $-$    &   $-$ &  $-$ &      $-$    &       $-$     \\
\hline
\end{tabular}
\end{center}
\end{table*}

%\bsp	
\label{lastpage}
\end{document}